\documentclass[12pt]{article}
\usepackage{authblk}
\usepackage{amsmath, amsthm, amssymb, appendix, amsfonts, comment, appendix}
\usepackage{multirow, array, dcolumn, lipsum}
\usepackage[export]{adjustbox}[2011/08/13]
\usepackage{hyperref}
\usepackage{mathrsfs}
\usepackage{graphicx}
\usepackage{subcaption}
\usepackage[dvipsnames]{xcolor}

\hypersetup{
linkcolor=blue,     
citecolor=blue,     
urlcolor=blue}      

\begin{document}
\title{Spherically Symmetric, Static Solutions in Presence of Matter-Curvature Coupling}
\author{Debanjan Debnath\thanks{debjanjan@gmail.com}}
\author{Kaushik Bhattacharya\thanks{kaushikb@iitk.ac.in}}
\affil{Department of Physics, Indian Institute of Technology Kanpur, Kalyanpur 208016, India}
\maketitle

\begin{abstract}
In this work we have proposed some spherically symmetric, static spacetimes in a theory of gravity which permits non-minimal coupling (NMC) between curvature of spacetime and fluid variables. It is shown that these non-minimally coupled theories may admit of new class of metric solutions. Known metric solutions from GR can also be solutions of the non-minimally coupled theories, for these cases the NMC affects the nature of the fluid which sources the spacetime. The paper presents multiple ways in which the modified field equations appearing in non-minimally coupled theories can be solved. The NMC produces multiple definitions of the stress-energy tensor. The paper discusses the complexity related to these sources of curvature as, unlike in minimally coupled general relativity, in the present theory the Ricci curvature itself can affect the stress-energy tensor of the effective fluid which seeds spacetime curvature. The various energy conditions related to various forms of possible stress-energy tensors are presented in the paper. 

\end{abstract}
\section{Introduction}

General relativity is by far, the most successful classical (non-quantum) theory describing the gravitational phenomena. Einstein was the first  to use general relativity to our universe \cite{Einstein1917} as a whole in the hope to grasp a rough picture of it. Hence, the history of modern cosmology is almost as old as general relativity. 

Incidentally, the Einstein's field equations can be  derived from, what is known as ``principle of least action" \cite{PoissonToolkit}, i.e., the general relativity admits a Lagrangian formulation \cite{PoissonToolkit, WaldGR}.  
The Lagrangian density ($\mathscr{L}$) or equivalently, the Lagrangian, for the gravitational field in vacuum was first given by Hilbert:
\begin{align}
    \mathcal{S}_{EH} = \frac{1}{2 \kappa} \int d^4x \sqrt{- g} \mathscr{L} = \frac{1}{2 \kappa} \int d^4 x  \sqrt{- g} R,
\end{align}
where $R$ is the Ricci scalar. It is then straightforward to vary this action with respect to the metric components $g^{\mu \nu}$ to get the field equations. 

This Lagrangian formulation helps us to generalize the theory and accommodate new kind of interactions in the gravitational sector in a quite general way. In the present work we will be using this formulation to discuss non-minimal coupling between the curvature of spacetime and the perfect fluid present in the spacetime. Such kind of interactions have been discussed previously in several contexts ranging from ``Modified Non-relativistic Dynamics" [MOND] \cite{bettoni1} to cosmological interests \cite{bettoni2}. Non-minimal coupling between curvature scalar and fluid \cite{goenner} is an esoteric topic which is interesting and perhaps very important because in the modern perspective of cosmology,  dark sector, comprising of the dark matter [DM] sector and the dark energy [DE] sector, is often described by an effective fluid under certain considerations \cite{FluidDE1, FluidDE2}. The dark  sector has a length scale related to it: the mean free path of the constituents of the dark sector [for discussion of relativistic fluid at various length scale, see \cite{RelFluidDy}]. For negligible  interaction between the constituent particles, this length scale is very large and consequently the dark sector deviates from the fluid approximation  which involves the transition from the particle limit to the  continuous limit. For length scales smaller than the cosmological scale, the dark sector is primarily dominated by the DM sector. It is inferred that, for self interacting dark matter models, this length scale can be in the astrophysical scales \cite{Spergel:1999mh, Tulin:2017ara}. This astrophysical scales are near to or greater than the galactic scales or scales corresponding to cluster of galaxies. It is interesting to note that in these cases there is a correspondence between the curvature scales of the spacetime and the length scales related to the dark sector when we model it as a fluid. One can expect that this correspondence may lead to a deeper connection between the gravitational sector and the fluid sector and these two sectors may actually affect each other resulting in a non-minimal coupling between them. In a similar way, in a larger length scale, which is nearer to the cosmological scale, one may expect the dark energy [DE] fluid to become the  determining factor in the dynamics. The length scale related to this fluid is around the cosmological scale. In this scales one can think of curvature and DE fluid couplings. In the cosmological backdrop, such couplings can be implemented in the Friedmann-Lemaitre-Robertson-Walker [FLRW] background models. In this work we will not study dynamic FLRW cosmology, rather we will be interested in the spherically symmetric, static solutions of modified field equations in presence of curvature-fluid coupling.  In some cases as discussed in Ref.~\cite{Debnath:2024urb}, one can argue that the end state of a gravitational collapse may contain some effects of the DE. In these cases the dark sector, comprising of interacting DM and DE fluids, can be modeled by an effective fluid which can produce spherically symmetric, static spacetimes in a length scale much smaller than the cosmological length scale where the effective fluid comprising the dark sector is coupled to the curvature of spacetime. There are many interesting ways one may end up with curvature-fluid coupled systems in various length scales going up from the galaxy cluster scale to the cosmological scale.  


One of the reasons why such non-minimal interactions between curvature and fluid variables must be studied seriously is because these interactions can always exist in principle. As because our knowledge of the dark sector is not that promising, all possibilities need be investigated properly and if we can rule out some possibilities, then we gain some important information in this process. Non-minimal coupling between fluids or fluid-scalar field interactions, primarily between DM and the dark energy (DE) sectors, are widely studied presently \cite{Debnath:2024urb, Saha:2024xbg, Hussain:2023kwk, Tamanini:2015iia, Boehmer:2015kta, Boehmer:2015sha}. In these papers the effect of curvature is not directly affecting the stress-energy tensor of the fluids or the scalar field, only the components of the dark sector interact with each other. In Ref.~\cite{Hussain:2023kwk} the authors proposed a non-minimal interaction between the dark sector components and curvature of spacetime. The primary aim of most of these works was to investigate how the non-minimal coupling [NMC] affects the late time accelerating behavior of the universe. In Ref.~\cite{Saha:2024xbg} the authors have investigated how NMC in the dark sector affects gravitational collapse in the late phase of the universe. All these works do show a considerable interest in NMC in cosmology. 

In Ref.~\cite{bettoni2} the authors studied NMC between curvature and fluid in the context of cosmology. In this article the authors discussed two particular kinds of NMC, they named these couplings a `Conformal coupling' and `disformal coupling'. The authors even presented a model of NMC  between DE and DM via curvature effects. One question naturally arises in this circumstance, if NMC between fluid variables and curvature can become relevant in the cosmological scale, can this kind of a NMC play any role in the cluster of galaxy scales? One can model galaxies and clusters of galaxies by some specific form of static, spherically symmetric solution of the Einstein equation. In Refs.~\cite{Bharadwaj:2003iw, Roberts:2002ei, Dey:2014gka, Dey:2013yga, Dey:2013vaa} the authors have proposed some forms of galactic spacetime, these are all spherically symmetric and static spacetimes. In Ref.~\cite{florides} the author has shown that there can be a spherically symmetric, static spacetime with uniform energy density. This spacetime is matched with an external Schwarzschild spacetime. This kind of a spaccetime can sometimes represent an Einstein cluster. From these works it becomes clear that many authors have attempted to study galactic or extragalactic physics by employing various kinds of spherically symmetric, static solutions of the Einstein equation. In all of the above-mentioned works we do not have any NMC between curvature and the fluid variables. In most of the cases where we deal with extra galactic spacetimes we have the presence of the dark sector and in this cases the next interesting step can be to calculate the fate of spherically symmetric, static solutions of the Einstein equation in presence of a non-minimal coupling between the fluid comprising the dark  sector and the curvature of spacetime.  Near the cosmological scale the dark sector is dominated by the dark energy component alone. In the present work we will see that one can obtain spherically symmetric, static spacetimes even in presence of pure cosmological constant. This is a new prediction and one can devise some physical probes to test the validity of this prediction. If correct, these kind of spacetimes will be the largest spherically symmetric, static structures in the universe. Most of the solutions presented in this paper all hint at curvature-fluid couplings between the Ricci scalar and some  negative pressure, dark sector fluid. 

In the present work, we propose various spherically symmetric, static spacetimes which can be the solutions of the modified Einstein equation when we take into account the curvature-fluid interaction at the level of the action. In problems related to NMC between curvature and fluid variables one encounters two unknown sectors: the spacetime metric and the form of the non-minimal coupling term. The modified Einstein equation can only shed light on one of these two unknowns when we apriori fix the form of the other unknown sector. In the present work we will solve the modified field equations in two different ways. In the first approach, we will assume a suitable form of the non-minimal coupling term, between curvature and fluid variables, and then proceed to find out the static, spherically symmetric spacetime metric resulting from such a non-minimal coupling by solving the modified field equation. This process gives rise to new class of spacetimes which may include some previously known solutions in general relativity [GR].  In the second approach, we start with a known spherically symmetric, static solution of the Einstein equation in GR and try to find out, under what conditions these spacetime can be a solution of the modified Einstein equation in presence of NMC between curvature and fluid variables. These conditions will then specify the  form of the NMC term. When this scheme of action predicts reasonable non-minimal coupling effects, we say that the problem is solved.  

We have given special attention to the energy conditions in the present case as whenever there is NMC involving the fluid variables, the energy condition of the fluid gets altered.  NMC between the fluid sectors or the fluid sector and curvature creates new kind of interpretational difficulties as because in these cases the fluid sectors are always interacting and their identity is in question. Due to this basic difficulty for non-minimally coupled fluids, we have defined multiple stress-energy tensors [SETs] for the fluid in question. Each proposed SET focuses on some different physical aspects, of the non-minimal interaction term, and define the energy density and pressure of the fluid in different ways. This interpretational part is important for our work because if the energy density of the fluid corresponds to some dark sector energy density then one has to properly choose a particular SET out of the various possible ones. We have tried to answer the question in our way in this paper.  

Before we present the main calculations we would like to specify our conventions regarding the unit system and spacetime metric. The signature of the metric we are working with is $(-, +, +, +)$. We also set the units such that 
\begin{align}
    G = \frac{1}{8\pi}, \,\,\,\,\ c = 1, \,\,\,\,\ \frac{h}{2\pi} = 1, \,\,\ \text{and} \,\,\,\,\ k_B = 1;
\end{align}
where $G$, $c$, $h$ and $k_B$ are Newton's gravitational constant, the speed of light in vacuum, the Planck's constant and Boltzmann's constant respectively. The superscripts [or subscripts] $0, 1, 2, 3$ on any quantity stand for the components $t, r, \vartheta, \phi$ respectively. Greek indices, e.g., $\mu$, $\nu$ etc., will assume values from $0$  to $3$ and the Latin indices, e.g., $i$, $j$ etc., will take values $1$ to $3$ and will be designated as ``spatial components". A prime [or a dot] over any quantity signifies a partial differentiation [of that quantity] with respect to the radial coordinate $r$ [or the temporal coordinate $t$]. Primes [or dots] with higher number are defined accordingly.  With these choice for units:
\begin{align}
    M^2_{Pl} \equiv \frac{c \hbar}{8 \pi G} = 1\,,\,\,\,\,\,\kappa \equiv \frac{8 \pi G}{c^4} = 1\,.
\end{align}
As for the definitions of the Christoffel connections and Riemann tensor, we adopt the following conventions:
\begin{align}
    \Gamma^{\lambda}_{\mu \nu} = \frac{1}{2} g^{\lambda \sigma} \left( \partial_{\mu} g_{\nu \sigma} + \partial_{\nu} g_{\sigma \mu} - \partial_{\sigma} g_{\mu \nu} \right), \,\,\, 
    &R^{\rho}_{\,\,\,\sigma \mu \nu} = \partial_{\mu} \Gamma^{\rho}_{\nu \sigma} - \partial_{\nu} \Gamma^{\rho}_{\mu \sigma} + \Gamma^{\rho}_{\mu \lambda} \Gamma^{\lambda}_{\nu \sigma} - \Gamma^{\rho}_{\nu \lambda} \Gamma^{\lambda}_{\mu \sigma}; \nonumber \\
   \text{and}, \,\, R_{\mu \nu} &= R^{\lambda}_{\,\,\,\mu \lambda \nu}, \,\,\, R = g^{\mu \nu}  R_{\mu \nu}.
\nonumber   
\end{align}
The definition of the covariant derivative of a contravariant vector $A^{\mu}$ is given by: $\nabla_{\nu} A^{\mu} = \partial_{\nu} A^{\mu} + \Gamma^{\mu}_{\nu \lambda} A^{\lambda}.$
We will use the above conventions throughout the present paper.

This paper is organized as follows: in section \ref{FE&SETs} we discuss the action functional corresponding to a fluid which is in the presence of curvature coupling and the field equation follows from it. We then formulate various alternative ways of defining the stress-energy tensors. Then in section \ref{givenfc}, we present resultant spacetimes which we get for a assumed form of the conformal coupling function. In section \ref{unknfc} we follow the reverse procedure and find out the conformal coupling functions for assumed forms of the spacetime metric. Then in section \ref{econds} we discuss at length, the familiar energy conditions which the matter-sector is supposed to  follow in various spacetimes we have obtained. Finally in section \ref{conclusion}, we conclude the paper by discussing various technical issues which arises in the scheme which is presented.

\section{Conformally Coupled Perfect Fluids and the Modified Field Equation} \label{FE&SETs}

In Ref.~\cite{bettoni2} the authors hinted at two different kinds of fluid-curvature couplings. In the present article we will primarily focus on one of the types of fluid-curvature couplings where the  fluid variables directly couples with the Ricci scalar of the spacetime at the level of the action. This kind of fluid-curvature NMC was named by the previous authors as the conformal coupling and we will follow their terminology and call the NMC as conformal coupling. In this paper we do not present the results related to the other kind of fluid-curvature coupling: The disformal coupling. We will work with disformal coupling in a separate publication as it involves coupling which is more intricate in nature and the problem related to it requires different techniques to solve.

\subsection{Field Equation in Conformal Coupling} 

The action of a perfect fluid in general curved space which is non-minimally coupled with the Ricci scalar can be written as
\begin{align}
    \mathcal{S}_c = \frac{1}{2 \kappa} \int d^4x \sqrt{-g} \hspace{1mm}\left[1 + \alpha_c F_c(n, s) \right] R + \mathcal{S}_{fluid}.
\label{beqnnm}
\end{align}
Here $F_c$ is a function of the particle number density $n$ and entropy per particle $s$; $\alpha_c$ is the coupling constant, $R$ is the Ricci scalar and $\mathcal{S}_{fluid}$ is the action functional for the fluid and given by
\begin{align}
  \mathcal{S}_{fluid} = \int d^4x \sqrt{-g} \hspace{1mm} \left[F(n, s) + J^{\mu} (\psi_{, \hspace{0.3mm} \mu} + s \theta_{, \hspace{0.3mm} \mu} + \beta_A \alpha^{A}_{\hspace{1mm}, \hspace{0.3mm} \mu}) \right]. \label{Sfluid}
\end{align}
In the above, $F$ is also a function of the particle number density   of the constituent matter-fluid $n$,  entropy per particle $s$ and $J^{\mu}$ is the densitized particle number flux vector $\sqrt{-g} n U^{\mu}$, $U^{\mu}$ is the 4-velocity of a fluid particle and is normalized as $U^{\mu}U_{\mu} = -1$. The action is a functional of the metric $g_{\mu\nu}$, $J^{\mu}$, the Lagrangian coordinates  $\alpha^A$'s and the spacetime-scalars (or Lagrangian multipliers) $\psi$, $\theta$ and $\beta_A$. These Lagrangian multipliers are there to make the fluid satisfy various constraint equations, like, conservation of particle number density, i.e., $J^{\mu}_{\,\ , \hspace{0.3mm} \mu} = 0$, entropy exchange constraint, i.e., $\left(sJ^{\mu} \right)_{, \hspace{0.3mm} \mu} = 0$ etc. 

When we vary this action with respect to the metric components $g_{\mu \nu}$, we get the following field equation \cite{bettoni2}:
\begin{align}
    \frac{1}{\kappa} (1 + \alpha_c F_c) G_{\mu \nu} = g_{\mu \nu} F - h_{\mu \nu} \left(\frac{\partial F}{\partial n} + \frac{\alpha_c}{2\kappa}  \frac{\partial F_c}{\partial n} R \right) n - \frac{\alpha_c}{\kappa} \left( g_{\mu \nu} \Box F_c - \nabla_{\mu} \nabla_{\nu} F_c \right), \label{f_e}
\end{align}
where $h_{\mu \nu} = g_{\mu \nu} + U_{\mu} U_{\nu}$ is the transverse metric, $G_{\mu \nu}$ [$\equiv R_{\mu \nu} - \left(1/2 \right) g_{\mu \nu} R$] is the Einstein tensor and the operator, $\Box$, is defined as: $\Box \equiv g^{\mu \nu} \nabla_{\mu} \nabla_{\nu} $.

In the presence of non-minimal coupling,  where there are terms involving the coupling between the fluid variables and the second derivative of the metric, there exists multiple definitions of the stress-energy tensor.  All the different definitions give different physical interpretations about the system. 
Consequently one can define the energy density and pressures of the system in multiple ways.  Broadly, we can pursue two different approaches to define the energy density and pressure of the system which we discuss below.
\begin{enumerate}
    
\item We can define an effective gravitational constant denoted by $\kappa_{\text{eff}}$ and an effective stress-energy tensor.
    In this approach, we write field equations in Eq.~\eqref{f_e} as 
\begin{align}
    \frac{1}{\kappa_{\text{eff}}}  G_{\mu \nu} =  T^{\left(\text{eff}\right)}_{\mu \nu}, \label{f_e_1}
\end{align}
with
\begin{align}
   T^{\left( \text{eff} \right)}_{\mu \nu} &\equiv g_{\mu \nu} F - h_{\mu \nu} \left(\frac{\partial F}{\partial n} + \frac{\alpha_c}{2\kappa}  \frac{\partial F_c}{\partial n} R \right) n - \frac{\alpha_c}{\kappa} \left( g_{\mu \nu} \Box F_c - \nabla_{\mu} \nabla_{\nu} F_c \right), \\ \text{and} \,\ \kappa_{\text{eff}} &\equiv \kappa \left(1 + \alpha_c F_c \right)^{-1}.
\end{align}
It can be noted, that, the above stress-energy tensor can be written as a stress-energy tensor for an ideal fluid, plus some additional terms coming from the curvature coupling, i.e., 
\begin{align}
     T^{\left( \text{eff} \right)}_{\mu \nu} \equiv T^{\left( \text{fluid} \right)}_{\mu \nu} + T^{\left( \text{non-min} \right)}_{\mu \nu},
\end{align}
where
\begin{align}
    T^{\left( \text{fluid} \right)}_{\mu \nu} &\equiv p_{\text{fluid}} g_{\mu \nu} + \left( \rho_{\text{fluid}} + p_{\text{fluid}}  \right) U_{\mu} U_{\nu}, \label{SET_fluid} \\ \text{and},  \,\ T^{\left( \text{non-min} \right)}_{\mu \nu} &\equiv -  \left(\frac{\alpha_c}{2\kappa}  n \frac{\partial F_c}{\partial n}  R\right) h_{\mu \nu}  - \frac{\alpha_c}{\kappa} \left( g_{\mu \nu} \Box F_c - \nabla_{\mu} \nabla_{\nu} F_c \right). \label{SET_nonmin}
\end{align}
In the first of the above relations, we have defined
\begin{align}
    \rho_{\text{fluid}} \equiv - F, \,\,\ \text{and}, \,\,\ p_{\text{fluid}} \equiv F - n \frac{\partial F}{\partial n}. \label{1stpov}
\end{align}
We have to keep in mind that, this splitting of the $T^{(\text{eff})}_{\mu\nu} $ into a fluid part and a non-minimal part is a hypothetical one. The stress-energy tensor corresponding to the fluid  part is not conserved, $\nabla^{\mu}\left(\kappa_{\text{eff}}\,T^{(\text{fluid})}_{\mu \nu}\right) \ne 0$, although $\nabla^{\mu}\left(\kappa_{\text{eff}}\,T^{(\text{eff})}_{\mu \nu}\right) = 0$. 

The stress-energy tensor (SET) $T^{(\text{fluid})}_{\mu \nu}$ of the fluid part in the present case coincides with the standard SET for an ideal fluid in GR. The only difference of this SET from the analogous SET of GR is manifested in the non-conservation of $T^{(\text{fluid})}_{\mu \nu}$ in the present case. As there is no standard observational method to find the components of $T^{\left( \text{non-min} \right)}_{\mu \nu}$, we do not directly physically observe the components of $T^{(\text{eff})}_{\mu \nu}$.  
    
\item Alternatively, in another way the total energy density and pressure for this system can be reasonably defined from the Euler's equations which can be derived from the field equations, Eq.~\eqref{f_e}, and comparing those with the corresponding equations obtained from Einstein's field equations for minimally coupled matter. This is the route in which the previous authors defined the total energy density and pressure in presence of NMC in Ref.~\cite{bettoni2}. Acting with the covariant derivative $\nabla^{\mu}$, on Eq.~\eqref{f_e} and taking the inner product of resultant equation with $U^{\nu}$ and $h^{\mu \nu}$ respectively [i.e., along and orthogonal to $U^\mu$] we get
\begin{align}
    \left(\frac{\partial F}{\partial n} + \frac{\alpha_c}{2 \kappa} R \frac{\partial F_c}{\partial n}  \right) \left( \dot{n} + \theta n  \right) + \left(\frac{\partial F}{\partial s} + \frac{\alpha_c}{2 \kappa} R \frac{\partial F_c}{\partial s}  \right) \dot{s} = 0,  \\
    n \left(\frac{\partial F}{\partial n} + \frac{\alpha_c}{2 \kappa} R \frac{\partial F_c}{\partial n}  \right) \dot{U}^\mu - h^{\mu \nu} \nabla_{\nu} \left[F - n\frac{\partial F}{\partial n} + \frac{\alpha_c}{2 \kappa} R  \left(F_c - n\frac{\partial F_c}{\partial n}  \right) \right] + \frac{\alpha_c}{2 \kappa} F_c \, h^{\mu \nu} \nabla_\nu R = 0. \label{EoMEulNonmin}
\end{align}
The second equation corresponds to the analogous Euler's equation in GR with a source term which is induced by the NMC. 
Here we have denoted $\nabla_\mu U^\mu$ by $\theta$ and the dot over a  quantity denotes the operation of acting $U^\mu \nabla_\mu$ on that quantity. 
Comparing these with the general relativistic analogues (with the source term in the Euler's equation)
\begin{align}
   \frac{\partial \rho}{\partial n} \left( \dot{n} + \theta n  \right) +\frac{\partial \rho}{\partial s} \dot{s} = 0, \,\,\, \left(\rho + p \right) \dot{U}^\mu + h^{\mu}_{\nu} \nabla^{\nu} p  + \frac{\alpha_c}{2 \kappa} F_c \, h^{\mu \nu} \nabla_\nu R= 0,
\end{align}
compels us to define 
\begin{align}
    \rho_{\text{Eul}} \equiv - F - \frac{\alpha_c}{2\kappa} R F_c \,\,\,\ \text{and} \,\,\,\ p_{\text{Euler}} \equiv n \frac{\partial \rho_{\text{Eul}}}{\partial n} - \rho_{\text{Eul}}. \label{p_rho_tot_def} 
\end{align}
In the above, an additional negative sign is introduced to make it consistent with the minimal case when $\alpha_c$ is equal to zero and the label ``Eul" stands for the fact that, the energy density and pressure thus defined, have its origin from the analogous consideration of Euler's fluid equation in minimally coupled theory.
The total pressure can be rewritten as
\begin{align}
    p_{\text{Eul}} = - \left(n \frac{\partial F}{\partial n} - F\right) - \frac{\alpha_c}{2\kappa} \left(n \frac{\partial F_c}{\partial n} - F_c \right) R.
\end{align}
The first term in the above expression corresponds to the minimally coupled sector and the second term arises because of the non-minimally coupled sector. When the coupling constant $\alpha_c$ vanishes we get back our previous expressions for the energy density and pressure in minimally coupled case. It can be seen, that, in this approach, the total energy density (and the pressure) get it's contribution also from the curvature coupling.

It can be seen, that, if we construct a stress-energy tensor $T^{(\text{Eul})}_{\mu\nu}$ corresponding to a perfect fluid which has $\rho_{\text{Euler}}$ and $p_{\text{Eul}}$ as energy density and pressure, then
\begin{align}
    T^{(\text{Eul})}_{\mu\nu} = p_{\text{Eul}} g_{\mu\nu} + \left(\rho_{\text{Eul}} + p_{\text{Eul}} \right) U_{\mu} U_{\nu},
\end{align}
then using Eq.~\eqref{p_rho_tot_def}, we can check that
\begin{align}
    T^{(\text{Eul})}_{\mu\nu} = g_{\mu \nu} F - h_{\mu \nu} \left(\frac{\partial F}{\partial n} + \frac{\alpha_c}{2\kappa}   \frac{\partial F_c}{\partial n} R \right)n.
\end{align}
Consequently $T^{(\text{Eul}}_{\mu \nu} $ is also not conserved $\nabla^{\mu} T^{(\text{Eul})}_{\mu \nu} \neq 0.$

\end{enumerate}
As $p_{\text{Eul}}$ and $\rho_{\text{Eul}}$ are directly related to relativistic fluid flow, we will assume that these quantities to be physically observable if one can figure out the flow pattern. The properties of the relativistic fluid in a NMC system can yield physically measurable definitions of pressure and energy density. It is to be noted that even in static situations, $p_{\text{Eul}}, \rho_{\text{Eul}}$ differ from  $p_{\text{fluid}}, \rho_{\text{fluid}}$ for cases where $R\ne 0$ and $\alpha_c \ne 0$. This implies that the effective fluid flow in a NMC system is affected by the nature of the curvature-fluid coupling. This feature can have interesting outcome in a NMC system. One can have cases where the non-minimal coupling term can nullify the effect of the fluid energy density and consequently produce a system where $\rho_{\text{Eul}}=0$ although $\rho_{\text{fluid}} >0$. Such zero effective energy density systems do occur in bouncing cosmologies in spatially flat Friedmann-Lemaitre-Robertson-Walker background \cite{Battefeld:2014uga}, where exactly at the cosmological bounce the effective energy density of the system becomes zero (due to the presence of some exotic matter component) forcing the Hubble parameter to vanish momentarily. In the present case, as we are working with static solutions, once the effective energy density vanishes, it remains so throughout. We will encounter one such case, where we can have a solution of the NMC theory, where  $\rho_{\text{Eul}}$ vanishes.   

From the above discussion it is clear that in the presence of fluid-curvature coupling one can define multiple forms of stress-energy tensor (SET). One can also define a third kind of SET which does not differ in their component values from the SET in GR and consequently we do not discuss it here. We briefly opine about this third form of the SET in the concluding remarks.   

For simplicity, we will consider only static and spherically symmetric spacetime having spacetime interval (will also be mentioned as metric or metric components for brevity) of the following form:
\begin{align}
    ds^2 = - e^{2\alpha(r)} dt^2 + e^{2\beta(r)} dr^2 + r^2 d \Omega^2. \label{SphSymmLine}
\end{align}
In the above expression, we have defined
\begin{align}
    d \Omega^2 \equiv d\vartheta^2 + \sin^2\vartheta d\phi^2.
\end{align}
The expressions for the Christoffel symbols, Ricci scalar and other identities which we will use in the following text are given in Appendix A.

\section{Solution of the Field Equations for a Given Conformal Coupling Function $\boldsymbol{F_c(n,s)}$}
\label{givenfc}

In this section we will solve the modified field equations, Eq.~(\ref{f_e}), by assuming some simple forms of the conformal coupling function $F_c(n,s)$. The solution will yield a spacetime metric and equation of state (EoS) of the fluid present in the spacetime. Initially we assume that $F_c$ is given as a function of $r$, later we find out the $n,s$ dependence of $F_c$ after we get the functional dependence of $n$ on the radial coordinate. In the static cases we will in general assume $F_c(n)$, here the entropy density will not play any active role, as we do not assume any heat transfer between various parts of the fluid, and hence we will assume all the fluid variables to be purely functions of $n$.  In this case the field equations in Eq.~(\ref{f_e}) yield:
\begin{align}
    \left(1 + \alpha_c F_c \right) G^0_0 &= F  - \alpha_c e^{-2\beta} \left[ {F_c}^{\prime \prime} - \left(\beta^{\prime} - \frac{2}{r} \right) {F_c}^{\prime} \right], \label{FE00_C} \\
   \left(1 + \alpha_c F_c \right) G^1_1 &= F  - n \frac{\partial F}{\partial n} - \frac{\alpha_c }{2} n \frac{\partial F_c}{\partial n} R - \alpha_c  e^{-2\beta} \left(\alpha^{\prime} + \frac{2}{r} \right) {F_c}^{\prime},  \label{FE11_C} \\
   \left(1 + \alpha_c F_c \right) G^2_2 &= F  - n \frac{\partial F}{\partial n} -  \frac{\alpha_c }{2} n \frac{\partial F_c}{\partial n} R  -  \alpha_c e^{-2\beta} \left[ {F_c}^{\prime \prime} + \left(\alpha^{\prime} - \beta^{\prime} + \frac{1}{r} \right) {F_c}^{\prime} \right]. \label{FE22_C} 
\end{align}
Subtracting Eq.~\eqref{FE22_C} from Eq.~\eqref{FE11_C} we get 
\begin{align}
       \left(1 + \alpha_c F_c \right) \left(G^1_1 - G^2_2 \right) =   \alpha_c e^{-2\beta} \left[ {F_c}^{\prime \prime} - \left(\beta^{\prime} + \frac{1}{r} \right) {F_c}^{\prime} \right].
\end{align}
Using the expressions for the Einstein tensor we can rewrite the above equation as
\begin{align}
    \left(1 + \alpha_c F_c \right)\left( \alpha^{\prime \prime} + {\alpha^{\prime}}^2 - \alpha^{\prime} \beta^{\prime} - \frac{\alpha^{\prime} + \beta^{\prime}}{r} - \frac{1 - e^{2 \beta} }{r^2} \right) + \alpha_c  \left[ {F_c}^{\prime \prime} - \left(\beta^{\prime} + \frac{1}{r} \right) {F_c}^{\prime} \right] =  0. \label{diff_metcomp}
\end{align}
This is a coupled differential equation which cannot be solved unless we know the functional form of $F_c$ and the functional form of $\alpha$ or $\beta$. If we assume a functional form of $\beta$, then the above equation becomes a differential equation for $\alpha$ for a given $F_c (r)$. Here we will solve the above differential equation for the simplest possible scenario, when, $\beta$ is a constant.

\subsection{The form of the emerging spacetime when $\boldsymbol{\beta(r)}$ is a constant}
\label{c0cn0}

In the present case, if we assume that $\beta$ is a constant, then,  Eq.~\eqref{diff_metcomp} simplifies to: 
\begin{align}
     \alpha^{\prime \prime} + {\alpha^{\prime}}^2  - \frac{\alpha^{\prime}}{r} - \frac{1- e^{2\beta}}{r^2}  + \alpha_c \left(1 + \alpha_c F_c \right)^{-1} \left( {F_c}^{\prime \prime} -  \frac{1}{r}  {F_c}^{\prime} \right) =  0. \label{DiffEqAlpha}
\end{align}
To solve the above differential equation for $\alpha(r)$ we have to assume a functional form of $F_c$. We assume the following power law form:
\begin{align}
    F_c (r) =  k \, r^{\xi}, \label{Fc2jmn}
\end{align}
where $k$ and $\xi$ are constants. By convention, we will assume that the conformal function $F_c$ is dimensionless, and consequently, the dimension of the constant $k$ becomes $L^{-\xi}$. Substituting this in Eq.~\eqref{DiffEqAlpha} we get
\begin{align}
     \alpha^{\prime \prime} + {\alpha^{\prime}}^2  - \frac{\alpha^{\prime}}{r} - \frac{1- e^{2\beta}}{r^2}  +  \frac{ \alpha_c k \, \xi \left(\xi - 2 \right)  r^{\xi -2} }{1 + \alpha_c  k \, r^{\xi}} = 0.
\end{align}
For simplicity we further impose $\xi = 2$, for which, the last term in the  above differential equation vanishes and we get the following solution:
\begin{align}
  \alpha (r) =  \ln \left(\frac{r}{r_b} \right) - \sqrt{2 - e^{2 \beta}} \ln \left(\frac{r}{r_b} \right) + \ln \left[ \left(\frac{r}{r_b} \right)^{2 \sqrt{2 -e^{2\beta}}} + C \right], \label{alphasol}
\end{align}
where $r_b$ and $C$ are integration constants. We will now discuss the nature of this solution in two different cases where the individual cases depend on the value of the constant $C$. 

\subsubsection{Case I: when $\boldsymbol{C = 0}$}

If we assume that $C = 0$, then from the above relation we can write
\begin{align}
    e^{2\alpha} =  \left(\frac{r}{r_b} \right)^{2 \widetilde{\beta} }, \,\,\, \text{with} \,\,\, \widetilde{\beta} \equiv  \left(1 + \sqrt{2 - e^{2 \beta}} \right).
\end{align}
From the above definition of $\widetilde{\beta}$, we can see that, we need 
\begin{align}
    e^{- 2 \beta} \geq \frac{1}{2}, \label{betacondCzero}
\end{align}
to make it a real number. The non-vanishing components of Einstein tensor, for this metric solution, are computed and given in Eq.~\eqref{ET_reverse_metric_Czero} of Appendix A. We can rewrite Eq.~\eqref{FE00_C} as
\begin{align}
       F  = \left(1 + \alpha_c F_c \right) G^0_0 + \alpha_c e^{-2\beta} \left[ {F_c}^{\prime \prime} - \left(\beta^{\prime} - \frac{2}{r} \right) {F_c}^{\prime} \right],
\end{align}
which yields, after substituting the expressions of $G^0_0 $ and $F_c \,$, the following:
\begin{align}
    F (r) = \frac{e^{-2\beta}-1}{r^2} + \alpha_c k \left(7 e^{-2\beta}-1\right). \label{Fsoljmn2}
\end{align}
From Eq.~\eqref{ricci_scalar_gen}, in the Appendix, we can calculate the Ricci scalar associated with this spacetime:
\begin{align}
    R =  \frac{R_{0}}{r^2}, \,\, \text{with} \,\, R_{0} \equiv 2 \left[1 - e^{-2\beta} \left(1 + \widetilde{\beta} + {\widetilde{\beta}}^2 \right) \right]. \label{ricciCzero}
\end{align}
To calculate the expression for the particle number density we rewrite Eq.~\eqref{FE11_C} as
\begin{align}
      \left(1 + \alpha_c F_c \right) G^1_1 &= F  - \frac{n}{n^{\prime}} \left( F^{\prime} + \frac{\alpha_c }{2}   R {F_c}^{\prime}  \right) - \alpha_c  e^{-2\beta} \left(\alpha^{\prime} + \frac{2}{r} \right) {F_c}^{\prime}.
\end{align}
Using the expressions for $\alpha (r)$, $G^1_1 (r)$, $F (r)$ and  $F_c (r)$, the above equation becomes
\begin{align}
    \frac{1}{\sqrt{\left|k\right|}} \frac{n^{\prime}}{n} = \frac{\widetilde{\mathcal{N}_1}  + \widetilde{\mathcal{N}_2} \widetilde{r}^2}{ {\mathcal{N}_1} \widetilde{r} +  {\mathcal{N}_2} \widetilde{r}^3}, \label{n_diffeq_reverse} 
\end{align}
where we have defined a dimensionless variable $\widetilde{r}$ as
\begin{align}
    \widetilde{r} \equiv r \sqrt{\left|k \right|} ,
\end{align}
and the expressions for the dimensionless constants $\widetilde{\mathcal{N}_1}$,  $\widetilde{\mathcal{N}_2}$, $\mathcal{N}_1$ and  $\mathcal{N}_2$ are given as follows:
\begin{align}
    \widetilde{\mathcal{N}_1} \equiv 2 \left(1 - e^{- 2\beta} \right), \,\,\, 
    \widetilde{\mathcal{N}_2} \equiv \alpha_c R_0,  \,\,\,
    \mathcal{N}_1 \equiv - 2\widetilde{\beta} e^{-2\beta}, \,\,\,
    \mathcal{N}_2 \equiv 2 \alpha_c  e^{- 2\beta} \left(1 - 2 \widetilde{\beta} \right). \label{Ns}
\end{align}
The above differential equation has the following solution:
\begin{align}
    n (\widetilde{r}) = n_0 \widetilde{r}^{\frac{\widetilde{\mathcal{N}_1}}{\mathcal{N}_1}} \left( \mathcal{N}_1 + \mathcal{N}_2 \widetilde{r}^2\right)^{\frac{1}{2} \left(\frac{\widetilde{\mathcal{N}_2}}{\mathcal{N}_2} - \frac{\widetilde{\mathcal{N}_1}}{\mathcal{N}_1} \right)}. \label{nsol1}
\end{align}
In the above expression, $n_0$ is an integration constant whose sign has to be chosen such that, the overall sign of $n$ comes out to be positive. This equation is too complicated to invert and write $r$ as a function of $n$. To make it simple, we will consider a special case where we take
\begin{align}
    e^{- 2 \beta} = 1, \,\, \text{and} \,\,\, \alpha_c k > 0\,,
    \label{expminustwobetaone}
\end{align}
which is consistent with Eq.~\eqref{betacondCzero}, and consequently
\begin{align}
   \widetilde{\beta} = 2, \,\,\,  \widetilde{\mathcal{N}_1} = 0, \,\,\, 
    \widetilde{\mathcal{N}_2} = - 12 \alpha_c,  \,\,\,
    \mathcal{N}_1 = - 4 \widetilde{\beta}, \,\,\,
    \mathcal{N}_2 = - 6 \alpha_c; \label{Ns2}
\end{align}
and consequently Eq.~\eqref{nsol1} becomes
\begin{align}
    n (\widetilde{r}) =  - 2 {n_0} \left(4 + 3 \left|\alpha_c \right| \widetilde{r}^2 \right). \label{n_soln_1_reverse}
\end{align}
As it was mentioned before, in order to make the particle number density a positive number, we have to assume $n_0$ a negative number.
The above equation [Eq.~\eqref{n_soln_1_reverse}] can be inverted to write $r$, as a function of particle number density $n$ as
\begin{align}
    \widetilde{r} = \sqrt{\frac{1}{3 \left|\alpha_c \right|} \left( \frac{n}{2 \left|n_0 \right|} - 4\right)} \,. 
\end{align}
In this case, the rescaled radial coordinate $\widetilde{r}$ varies from $0$ to $+\infty$, where $n$ varies from the value $8 \left|n_0 \right|$ at the origin and increases to $+\infty$ with $\widetilde{r}$.

However, when we take  $ \alpha_c k < 0$ [i.e., $\alpha_c k = - \left|\alpha_c k \right| = - \left|\alpha_c\right| \left|k\right|$ effectively], then, instead of Eq.~\eqref{n_soln_1_reverse}, we get
\begin{align}
    n \left( \widetilde{r} \right) = - 2 n_0 \left(4 - 3 \left|\alpha_c \right|{\widetilde{r}}^2 \right). \label{nsol_neg}
\end{align}
So, if we take the constant $n_0$ as a negative number, then, we can invert the above expression of $n (\widetilde{r})$ to write
\begin{align}
    \widetilde{r} = \sqrt{\frac{1}{3 \left|\alpha_c \right|} \left( 4-\frac{n}{2 \left|n_0 \right|}\right)} \,. 
\end{align}
From Eq.~\eqref{nsol_neg}, we see that, $n$ remains positive from the origin up to $\widetilde{r}_{\text{max}}$, where 
\begin{align}
    \widetilde{r}_{\text{max}} \equiv \sqrt{\frac{4}{3\left|\alpha_c \right|}},
\end{align}
and consequently our solution is valid also up to $\widetilde{r}_{\text{max}}$ in this case. From Eqs.~\eqref{Fc2jmn} and ~\eqref{Fsoljmn2} we can write the conformal functions in terms of the particle number density as follows:
\begin{align}
    F_c (n) =  \frac{1}{3 \left|\alpha_c \right|} \left( \frac{n}{4 \left|n_0 \right|} - 1\right), \,\,\,\,  F (n) = 6\alpha_c k.
\end{align}
The metric in this case becomes
\begin{align}
     ds^2 = -  \left(\frac{r}{r_b}\right)^{4} dt^2 +  {dr^2} + r^2 d \Omega^2.
     \label{reverse_metric}
\end{align}
From Eq.~\eqref{ricciCzero}, we can see that, this spacetime has a curvature singularity at $r=0$.

We can now calculate the energy density and pressure associated with the matter. In the first point of view,  we get [Eqs.~\eqref{1stpov}, ~\eqref{Fsoljmn2}, ~\eqref{n_diffeq_reverse} and ~\eqref{expminustwobetaone}]
\begin{align}
    \rho_{\text{fluid}} = - 6 \alpha_c k, \,\,\, 
    p_{\text{fluid}} = 6\alpha_c k. \label{rhopsolCzero}
\end{align}
To impose the positivity on the energy density, we assume [Eq.~\eqref{rhopsolCzero}]
\begin{align}
    \alpha_c k < 0.
\end{align}
As, for the chosen sign of  $\alpha_c k$, the pressure $p_{\text{fluid}}$ is a negative quantity. We can calculate the corresponding equation-of-state parameter (EoS), which we defined as
\begin{align}
    w_{\text{fluid}} \equiv \frac{p_{\text{fluid}}}{\rho_{\text{fluid}}}. \label{EoSfluid}
\end{align}
Using Eq.~\eqref{rhopsolCzero}, the above quantity becomes
\begin{align}
    w_{\text{fluid}} = - 1, \label{EoSexppov1Czero}
\end{align}
This result shows that pure dark energy like fluid, which is non-minimally coupled to curvature, can produce spherically symmetric, static spacetimes.


In the second point of view, using Eqs.~\eqref{Fsoljmn2}, ~\eqref{Fc2jmn}, ~\eqref{ricciCzero} and ~\eqref{n_diffeq_reverse},  we can write
\begin{align}
    \rho_{\text{Eul}} &=  -\frac{e^{-2\beta}-1}{r^2} - \alpha_c k e^{- 2\beta} \left(6 - \widetilde{\beta} - {\widetilde{\beta}}^2  \right), \label{rhoSolCzeroPov3}\\
    p_{\text{Eul}} &= \frac{\widetilde{P}_0 + \widetilde{P}_1 k \, r^2 + \widetilde{P}_2 \left(k \, r^2 \right)^2}{r^2 \left(P_0 + P_1 k \, r^2\right) } + \alpha_c k \left(7 e^{- 2 \beta} -1 \right). \label{pSolCzeroPov3}
\end{align}
The constants $\widetilde{P}$'s and $P$'s are given in Eq.~\eqref{P's} of Appendix. In this case  one can verify that $\rho_{\text{Eul}}=0$ if we assume the condition in Eq.~(\ref{expminustwobetaone}). The calculations show that although the fluid present in the spacetime acts like the cosmological constant the effective fluid which is physically perceived to take part in fluid motion has zero energy density and finite pressure which diverges at $r=0$ [Eq.~\eqref{pSolCzeroPov3}]. We infer that this system shows an effective EoS as $|w_{\text{Eul}}| \to \infty$ as we approach the origin. 

One must also note that the metric solution given in Eq.~(\ref{reverse_metric}) produces a value zero for the zero-zero component of the Einstein tensor. This fact shows that such a metric will not not satisfy standard energy conditions in GR and no standard fluid, even dark energy fluid, will be able to seed such a spacetime in minimally coupled GR. Only in the presence of curvature-fluid interaction, such a fluid can have dark energy like matter with positive energy density. Ultimately the curvature-fluid coupling suppresses the dark energy fluid density and consequently one gets a vanishing effective energy density for the effective fluid. 

In the present case we see that we can generate a spherically symmetric, static solution of the  field equations where the non-minimally coupled fluid present in the system acts like dark energy, in a limited region of space. In the present case we do not get dark matter like fluid. This kind of a solution requires to be matched with some other metric solution at $r=\widetilde{r}_{\text{max}}$. In this paper we do not discuss the matching conditions in NMC systems and hence we will not explicitly produce the whole solution. We note that at $r=\widetilde{r}_{\text{max}}$ we have $n(\widetilde{r})=0$ and consequently at this distance one can terminate the above kind of a solution. Physically the above solution looks like a blob of cosmological constant captured in a limited region of space. 
\subsubsection{Case II: when $\boldsymbol{C \neq 0}$}

It can be shown, that, when $C$ is non-vanishing, we  can identify the resultant spacetime with the JMN-2 spacetime, whose metric is given by \cite{Joshi2}
\begin{align}
    ds^2_{JMN-2} = -\frac{1}{16\lambda^2(2-\lambda^2)}\left[(1+\lambda)^2\left(\frac{r}{R_b}\right)^{1-\lambda}-(1-\lambda)^2\left(\frac{r}{R_b}\right)^{1+\lambda}\right]^2 &dt^2 \nonumber \\  + \left(2-\lambda^2 \right) \, &dr^2 + r^2d\Omega^2 \,;
 \label{JMN2metric}
\end{align}
where the parameter $\lambda$ has the following range: $0 \leq \lambda < 1$ and this spacetime is to be  matched with external Schwarzschild geometry at $r = R_b$. In other words, in GR, the JMN-2 spacetime has matter content from the origin [$r=0$], up to $r = R_b$. The metric in Eq.~\eqref{JMN2metric} has a curvature singularity at the origin, as can be shown by evaluating the Kretschmann scalar [see Eq. (43) of \cite{Joshi2}]. In the present case we have modified the field equations and the theory behaves differently from GR and consequently we do not interpret $R_b$ as one does in GR, here $R_b$ is a length scale which parametrizes the JMN-2 spacetime. In the following discussions, we consider the solution [Eq.~\eqref{alphasol}] with non-vanishing $C$, which can be rewritten as
\begin{align}
    e^{2 \alpha} = \left[C \left(\frac{r}{r_b}\right)^{1-\sqrt{2-e^{2 \beta}}} + \left(\frac{r}{r_b}\right)^{1+\sqrt{2 - e^{2 \beta}}}\right]^2.
\end{align}
We can match this with the time-time component of the metric [$g_{tt}$]  corresponding to the JMN-2 spacetime if we set
\begin{align}
     \frac{C}{r_b^{1 - \sqrt{2 - e^{2\beta}}}} = - \frac{(1 + \lambda)^2}{4 \lambda \sqrt{2 - \lambda^2}} \frac{1}{R^{1 - \lambda}_b}, \,\,\,
    &\frac{1}{r_b^{1 + \sqrt{2 - e^{2\beta}}}} =  \frac{(1 - \lambda)^2}{4 \lambda \sqrt{2 - \lambda^2}} \frac{1}{R^{1 + \lambda}_b}, \nonumber \\ 
   \text{and} \,\,\, \lambda = &\sqrt{2 - e^{2\beta}}.
\end{align}  
After an algebraic simplification we get the following two conditions for the two integration constants:
\begin{align}
    r_b = \left[ \frac{4 \lambda \sqrt{2 - \lambda^2}}{(1 - \lambda)^2} \right]^{\frac{1}{1 + \lambda}} R_b, \,\,\, C = - \left[16 \lambda^2 (1 - \lambda^2) \right]^{- \frac{\lambda}{1 + \lambda}}  (1 - \lambda)^{- 2  \left(\frac{1- \lambda}{1 + \lambda} \right)}  (1 + \lambda)^2.
\end{align}
It can also be seen, that, the metric component $g_{rr}$  becomes
\begin{align}
    e^{2 \beta} = 2 - \lambda^2,
\end{align}
which is automatically identical with the $g_{rr}$ component of the JMN-2 metric [Eq.~\eqref{JMN2metric}]. So we conclude that, the metric solution given by Eq.~\eqref{alphasol} with a proper choice for the constants $r_b$ and $C$, can be identified with the JMN-2 spacetime.
By substituting the expressions for $\alpha$ and $\beta$ in Eq.~\eqref{ricci_scalar_gen} from the Appendix, we can calculate the Ricci scalar, which can be written in the following form:
\begin{align}
   R = \frac{R_1 + R_2 r^{2 \lambda}}{r^2 \left(R_3 r^{2 \lambda} + R_4 \right)},
\end{align}
and the expressions for the constants $R_1$, $R_2$, $R_3$ and  $R_4$, are given as
\begin{align}
    R_1 =  &(1 + \lambda )^2 [1 + 3 (1 - \lambda ) \lambda ],  \,\,\,\, R_2 = - R^{-2\lambda}_b (1-\lambda)^2 [1 - 3 \lambda  (1 + \lambda )],  \,\,\,\,\nonumber \\ 
    &R_3  = - R^{- 2\lambda}_b \left(2 - \lambda^2 \right) (1 - \lambda)^2,  \,\,\,\, R_4 = \left(2 -\lambda^2 \right) (1 + \lambda)^2. 
    \nonumber
\end{align}
For future use, we also write the non-vanishing components of Einstein tensor in JMN-2 spacetime in Eq.~\eqref{ET_jmn2} of Appendix A.
From Eq.~\eqref{FE00_C} we can write:
\begin{align}
       F  = \left(1 + \alpha_c F_c \right) G^0_0 + \alpha_c e^{-2\beta} \left[ {F_c}^{\prime \prime} - \left(\beta^{\prime} - \frac{2}{r} \right) {F_c}^{\prime} \right],
\end{align}
which yields
\begin{align}
    F (r) = - \frac{1-\lambda ^2}{\left(2 - \lambda ^2 \right) \, r^2} + \frac{\alpha_c k \left(5 + \lambda^2 \right)}{2-\lambda^2}. \label{Fsol2}
\end{align}
To calculate the expression for the particle number density we rewrite Eq.~\eqref{FE11_C} as
\begin{align}
      \left(1 + \alpha_c F_c \right) G^1_1 &= F  - \frac{n}{n^{\prime}} \left( F^{\prime} + \frac{\alpha_c }{2}   R {F_c}^{\prime}  \right) - \alpha_c  e^{-2\beta} \left(\alpha^{\prime} + \frac{2}{r} \right) {F_c}^{\prime}.
       \nonumber
\end{align}
Using the expressions for $\alpha (r)$, $G^1_1$, $F (r)$ and  $F_c (r)$, the above equation yields a lengthy form of $n^\prime/n$.
The resulting equation cannot be integrated directly to obtain a solution in a closed form. What we can try, is to simplify it by a judicious choice of the constants and parameters $\alpha_c$, $k$, $R_b$ and $\lambda$. Assuming $\lambda = 1/2$, the form of the expression of $n^\prime/n$ simplifies and is given by:
\begin{align}
    \frac{n^\prime}{n} = \frac{54 - \frac{6}{R_b} r + 63 \alpha_c k r^2  + \frac{5 \alpha_c k}{R_b} r^3 }{-36 r+ \frac{12}{R_b} r^2 +\frac{16\alpha_c k}{R_b} r^4}. \label{ndiffeqJMN2}
\end{align}
We can integrate the above differential equation for some benchmark values of the dimensional constants, in our system of units,  as: $\alpha_c = 1$, $R_b = 1$ and $k = 1$. These assumed values of the dimensional constants are used to simplify the solutions and using these values we get the approximate solution:
\begin{align}
   \ln n (r)
   \approx \ln n_0 +  N_1 \log \left(N_2 - r \right)+ N_3 \log \left( r^2+ N_4 r+ N_5 \right) - N_6 \log r + N_7 \tan ^{-1}( N_8 r+ N_9);
\end{align}
where $n_0$ is the integration constant and the constants $N_1$ to $N_9$ are given by
\begin{align}
    N_1 = 1.64632, \,\,\, &N_2 = 1.12112, \,\,\, N_3 =0.08309, \,\,\, N_4 = 1.12112, \,\,\,  N_5 = 2.00692, \nonumber \\ 
    N_6 &= 1.5, \,\,\, N_7 = 1.67938, \,\,\, N_8 = 0.76862, \,\,\, N_9 = 0.43086.
    \nonumber
\end{align}
We rewrite the expression for $n (r)$ as:
\begin{align}
    n (r) \approx n_0 \left(N_2 - r \right)^{N_1} \left(r^2 + N_4 r + N_5 \right)^{N_3} r^{-N_6} e^{N_7 \tan^{-1} (N_8 r + N_9)}.
\end{align}
From the above expression of $n (r)$, it can be seen that, the particle number density diverges at the origin. This may be due to the  curvature singularity which JMN-2 spacetime has at the origin, as was mentioned previously. This expression for particle number density is complicated, and, cannot be easily inverted to write the radial coordinate $r$ as a function of $n$. However, near the origin, for small values of the radial coordinate [$r \ll N_9/N_8$], we can approximate the above relation as
\begin{align}
    n \approx \frac{\widetilde{n}_0}{r^{\frac{3}{2}}},
\end{align}
where we have defined $\widetilde{n}_0 \equiv n_0 N_2^{N_1} N_5^{N_3} e^{N_7 \tan^{-1} N_9}$. The above relation can be easily inverted.
From Eqs.~\eqref{Fsoljmn2} and ~\eqref{Fc2jmn} we can write the fluid-function $F$ and the conformal function $F_c$ as functions of the particle number density  for small radial distances as
\begin{align}
    F (n) \approx - \frac{3}{7} \left(\frac{n}{\widetilde{n}_0} \right)^{\frac{4}{3}} + 3, \,\,\, F_c (n) \approx \left(\frac{\widetilde{n}_0}{n} \right)^{\frac{4}{3}}. 
\end{align}
Utilizing the above results one can predict the various properties specifying the effective fluid present in the system.

We can now discuss about the energy densities and pressures resulting from various alternative points of view discussed before. Following the first point of view [Eq.~\eqref{1stpov}], and using Eq.~\eqref{Fsol2} we have
\begin{align}
   \rho_{\text{fluid}} = 3 \left( \frac{1}{7  r^2} - 1 \right), \label{rhosolpov1}
\end{align}
and the expression for $p_{\text{fluid}}$ can be obtained by using Eqs.~\eqref{n_diffeq_reverse2}, ~\eqref{rhosolpov1} and computing the following relation:
\begin{align}
    p_{\text{fluid}} = - \rho_{\text{fluid}} +  \left(\frac{n}{n^{\prime}} \right) \rho_{\text{fluid}}^\prime. \label{psolpov1}
\end{align}
This will be complicated expression and instead of writing the full expression, we consider to use it only to calculate it to calculate the EoS only for small values of the radial coordinate. From Eq.~\eqref{rhosolpov1}, we can see that, the fluid energy density $\rho_{\text{fluid}}$ is positive only within $0 < r \leq \frac{1}{\sqrt{7}}$. For small $r$, from Eqs.~\eqref{rhosolpov1} and ~\eqref{psolpov1} we get after a little algebra
\begin{align}
 \rho_{\text{fluid}} \approx \frac{3}{7 r^2}, \,\,\,   p_{\text{fluid}} \approx \frac{1}{7 r^2}.
\end{align}
From the above limiting values of the fluid energy density and pressure, the corresponding EoS [$w_{\text{fluid}}$] which we define as $\left({p_{\text{fluid}}}/{\rho_{\text{fluid}}}\right)$, seems to approach  $ \frac{1}{3}$, for a small value of the radial coordinate.

Using the prescription coming out from the Euler's equation, we can write the expression for the energy density as [Eq.~\eqref{p_rho_tot_def}]
\begin{align}
    \rho_{\text{Eul}} =  \frac{3}{7r^2} - \frac{441 - 37 r}{14 (9 - r) }, \label{rhosolnjmn2}
\end{align}
 and the expression for the pressure can be obtained by calculating
\begin{align}
    p_{\text{Eul}} = \left(\frac{n}{n^{\prime}} \right) \rho^\prime_{\text{Eul}} - \rho_{\text{Eul}}. \label{pjmn2}
\end{align}
 However, it can be seen from the form of the function $\rho_{\text{Eul}}$ [Eq.~\eqref{rhosolnjmn2}], that this solution diverges at $r = 9$ and also becomes negative after $r \approx 0.35$. As, in this case, it is problematic to define a positive valued energy density within a reasonable range of radial coordinate, we do not consider it further. 

One can note that the JMN-2 solution is a solution in GR, and there one can calculate
using Eq.~\eqref{ET_jmn2}, of the Appendix, that
\begin{align}
    \rho &= \left(\frac{1-\lambda^2}{2-\lambda^2}\right)\frac{1}{r^2}, 
\nonumber \\
    p &= \frac{1}{(2-\lambda^2)}\frac{1}{r^2}\left[\frac{(1-\lambda)^2 A - (1+\lambda)^2 B r^{2\lambda}}{A - B r^{2\lambda}}\right], 
\nonumber
\end{align}
where $A = \frac{(1+\lambda)^2 R_b^{\lambda-1}}{4\lambda \sqrt{2-\lambda^2}}$ and $B = \frac{(1-\lambda)^2 R_b^{-
\lambda-1}}{4\lambda \sqrt{2-\lambda^2}}$. For $\lambda = \frac{1}{2}$, the above expressions reduce to the following:
\begin{align}
    \rho = \frac{3}{7r^2}, \,\,\, 
    p  = \frac{9}{7r^2}\left(\frac{1-r}{9 - r}\right), \label{rhopsoleff}
\end{align}
which are well behaved functions for $0 < r < 9$.  There is a central singularity in JMN-2 spacetime and it occurs at $r=0$, on the other hand modified field equations produce another singularity at $r=9$, where the pressure diverges. The above results show that the values of $\rho_{\text{Eul}}, p_{\text{Eul}}$ or $\rho_{\text{fluid}}, p_{\text{fluid}}$ have some interesting similarities with their value  in GR. The results show how the NMC term changes the values of $\rho, p$ in JMN-2 spacetime.

Before we proceed to the next section we note that in this section we have shown how the field equations, in presence of curvature-fluid coupling, can be solved if we assume a form of the non-minimal coupling coupling function $F_c(n)$. Our method shows in the present case we can have new kinds of metric solutions. One of the solutions coming out from our analysis is actually the well known JMN-2 spacetime. We have explicitly shown when NMC can produce JMN-2 spacetime. It is observed from the expressions of the energy density and pressure of the effective fluid in JMN-2 spacetime that there is a singularity for a finite value of $r$ and consequently one must match the JMN-2 spacetime, with some other well behaved solution of the field equation, before this non-central singularity is attained. For systems where  fluid-curvature coupling is present one has to work out the matching conditions before JMN-2 spacetime can be matched to some external spacetime. Till now such matching conditions have not been worked out, we expect to work out matching conditions in non-minimally coupled scenarios in the near future.

We end this section by noting that, a similar analysis can be carried out by substituting $\alpha$ as a constant in  Eq.~\eqref{diff_metcomp}, and then solving the resultant differential equation for $\beta (r)$ for a suitably assumed form of the conformal function $F_c (r)$. However, we do not include the analysis in this paper.

\section{\fontsize{12}{15}\selectfont The Form of the Conformal Coupling Function $\boldsymbol{F_c(n)}$ for Spacetime Metrics Satisfying the Field Equations}
\label{unknfc}

In this section we will solve the field equation in a different way. In the present section, we will assume that the metric components are given functions of coordinates and then we will substitute these in the field equations and try to find out the unknown functions $F$ and $F_c$. For mathematical simplicity we will work with a spacetime whose metric components have the following generic form:
\begin{align}
    e^{- 2\beta(r)} = e^{2 \alpha(r)} = 1 + \frac{k}{r^m}, \label{metcompgenexam}
\end{align}
for some constant $k$ having the dimension of $L^m$ ($L$ stands for dimension of length). We substitute this in Eq.~\eqref{ricci_scalar_gen} appearing in the Appendix, to get
\begin{align}
    R = - \frac{k (m-1) (m-2)}{r^{m+2}}. \label{ricciscalargenexam}
\end{align}
Metric components, which produce the above form of the Ricci scalar, must be following the field equations 
which we write in the mixed components: 
\begin{align}
    \frac{1}{\kappa} (1 + \alpha_c F_c) G^{\mu}_{\nu} = \delta^{\mu}_{\nu} F - \left(\delta^{\mu}_{\nu} + U^{\mu} U_{\nu} \right) \left(\frac{\partial F}{\partial n} + \frac{\alpha_c}{2\kappa}  \frac{\partial F_c}{\partial n} R \right) n - \frac{\alpha_c}{\kappa} \left( \delta^{\mu}_{\nu} \Box F_c - \nabla^{\mu} \nabla_{\nu} F_c \right), \label{secFE}
\end{align}
which yields [from now on we will assume $\kappa = 1$ according to our convention and we can always revert back to the old convention by substituting $\alpha_c$ with $(\alpha_c/\kappa)$] the following three independent equations:
\begin{align}
    (1 + \alpha_c F_c) G^0_0 &=  F  - \alpha_c e^{2\alpha} \left[{F_c}^{\prime \prime} + {F_c}^{\prime} \left(\alpha^{\prime} + \frac{2}{r} \right) \right], \label{fe2_00} \\
   (1 + \alpha_c F_c) G^1_1 &=  F - \left(\frac{\partial F}{\partial n} + \frac{\alpha_c}{2}  \frac{\partial F_c}{\partial n} R \right) n  - \alpha_c e^{2\alpha} {F_c}^{\prime} \left(\alpha^{\prime} + \frac{2}{r} \right), \label{fe2_11} \\
   (1 + \alpha_c F_c) G^2_2 &=  F - \left(\frac{\partial F}{\partial n} + \frac{\alpha_c}{2}  \frac{\partial F_c}{\partial n} R \right) n  - \alpha_c e^{2\alpha} \left[{F_c}^{\prime \prime} + {F_c}^{\prime} \left(2\alpha^{\prime} + \frac{1}{r} \right) \right]. \label{fe2_22}
\end{align}
If we now subtract the second equation from the last one, we get
\begin{align}
       (1 + \alpha_c F_c) \left( G^2_2 - G^1_1 \right) =   - \alpha_c e^{2\alpha} \left[{F_c}^{\prime \prime} + {F_c}^{\prime} \left(\alpha^{\prime} - \frac{1}{r} \right) \right]. 
\end{align}
Now, unless $G^2_2$ and $G^1_1$ are equal, the left hand side does not vanish and the differential equation for $F_c$ contains $\alpha_c$. If we demand that the non-minimal interaction term will be first order in $\alpha_c$, as demanded by Eq.~(\ref{beqnnm}), then we have to assume that $F_c(n)$ is independent of the coupling constant.
If we want the expression for $F_c(n)$ to be independent of the coupling constant $\alpha_c$, then from Eq.~\eqref{ET_ex2} in the Appendix, equating the expressions for $G^1_1$ and $G^2_2$, we get the following solutions for $m$: $m = 1, -2$. We will consider only these two cases separately. In this section we only discuss the case related to $m=1$, the other case related to $m=-2$ does not turn out to be physically consistent, and we will briefly opine about it in the concluding section.

In this section we start to analyze the case when $m = 1$. In this case the metric components become
\begin{align}
    e^{- 2\beta(r)} = e^{2 \alpha(r)} = 1 + \frac{k}{r}. \label{met_form}
\end{align}
It can also be checked  [Eq.~\eqref{ET_ex2}] that the Einstein tensor vanishes
\begin{align}
   G_{\mu \nu} = 0 = G^{\mu}_{\nu} \,\,\,\,\ \text{for} \,\,\,\,\ \mu, \nu = 0, 1, 2, 3.
\end{align}
The same conclusion can be drawn by realizing that, this spacetime has a metric which has a form as Schwarzschild spacetime, which it turn, is a vacuum solution in Einstein gravity, which automatically implies the vanishing Einstein tensor.

Before we start solving the field equations, it should be mentioned at this point, that, the non-diagonal components of the field equations do not give any new equation. Both sides of the equations vanishes identically for the spacetime we have considered. To see this, let's consider the mixed components of the right hand side of the field equation in Eq.~(\ref{f_e}):
\begin{align}
    \delta^{\mu}_{\nu} F - \left(\delta^{\mu}_{\nu} + U^{\mu} U_{\nu} \right) \left(\frac{\partial F}{\partial n} + \frac{\alpha_c}{2}
    \frac{\partial F_c}{\partial n} R \right) n - \alpha_c
    \left( \delta^{\mu}_{\nu} \Box F_c - \nabla^{\mu} \nabla_{\nu} F_c \right)\,.
\end{align}
For non-diagonal components (i.e., $\mu \neq \nu$), the terms $\delta^{\mu}_{\nu} F$ and $\delta^{\mu}_{\nu} \Box F_c$ vanishes and as the fluid we are considering is static in the spacetime considered [Eq.~\eqref{SphSymmLine}], four-velocity has the components 
\begin{align}
    U^{\mu} \equiv (e^{-\alpha}, 0, 0, 0), \label{four_vel}
\end{align}
in our coordinate system, we also have $\delta^{\mu}_{\nu} + U^{\mu} U_{\nu} = 0$. The only term that remains to be checked to be zero is the last term of the above expression
\begin{align}
    \nabla^{\mu} \nabla_{\nu} F_c = g^{\mu \mu } \left(\frac{\partial^2 F_c}{\partial x^{\mu} \partial x^{\nu} } - \Gamma^r_{\mu \nu} {F_c}^{\prime} \right), \,\,\,\ \text{no sum over the index $\mu$}.
\end{align}
Remembering that $F_c$ is a function of $r$ only and using the form of the connection coefficients in Eq.~\eqref{connection_coeffs} of the Appendix, we can check that, for every non diagonal component each of the terms in the parenthesis vanishes separately. Hence it is verified that the non-diagonal components in the field equation do not produce extra equations other than the diagonal components and thus we will consider only the diagonal components. We should note that, this happens to be the case because $F_c$ is a function of the radial coordinate only.

In the present case all the independent field equations reduce to  the following form:
\begin{align}
    F  - \alpha_c \left(\Box F_c - \nabla^{t} \nabla_{t} F_c \right) = 0, \label{fe_00}\\
    F -  n \frac{\partial F}{\partial n} - \alpha_c  \left(\Box F_c - \nabla^{r} \nabla_{r} F_c \right) = 0, \label{fe_11} \\
    F - n \frac{\partial F}{\partial n} - \alpha_c  \left(\Box F_c - \nabla^{\vartheta} \nabla_{\vartheta} F_c \right) = 0. \label{fe_22}
\end{align}
From the last two equations we have the following constraint on the function $F_c$: $\nabla^r \nabla_r F_c = \nabla^{\vartheta} \nabla_{\vartheta} F_c$, giving  $g^{rr} \left(\partial^2_r {F_c} - \Gamma^r_{rr} \partial_r {F_c} \right) =  g^{\vartheta \vartheta}  \left(\partial^2_{\vartheta} F_c - \Gamma^r_{\vartheta \vartheta} \partial_r {F_c}\right)$, which implies
$$e^{-2\beta} \left({F_c}^{\prime \prime} - \beta^{\prime} {F_c}^{\prime} \right) =  \frac{1}{r} e^{- 2 \beta} {F_c}^{\prime}\,,$$
which can be rearranged to write the following differential equation:
\begin{align}
    {F_c}^{\prime \prime} - \left(\beta^{\prime} + \frac{1}{r}\right){F_c}^{\prime} = 0.
\end{align}
This equation can be integrated to solve for $F_c$:
\begin{align}
    {F_c}^{\prime} (r) = f_c r e^{
    \beta(r)} \implies F_c (r) = f_c \int dr \, r e^{\beta(r)} +  \widetilde{f_c} = f_c \int dr \, r \sqrt{\frac{r}{r + k}} + \widetilde{f_c}. \label{fc_soln}
\end{align}
where $f_c$ and $ \widetilde{f_c}$ are integration constants. The integration yields:
\begin{align}
     F_c(r) = \frac{1}{2} f_c \left(1 + \frac{k}{r} \right)^{1/2} \left[r^2 \left(1 - \frac{3k}{2r} \right) + \frac{3k^2}{2} \tanh^{-1}\left\{\left(1 + \frac{k}{r} \right)^{-1/2}\right\} \right] + \widetilde{f_c}. \label{fc(r)_soln}
\end{align}
If we substitute the expressions of ${F_c}^{\prime \prime}$ and ${F_c}^{\prime}$ from Eq.~\eqref{fc_soln} in the first of the field equations [Eq.~\eqref{fe_00}] and then using the Eqs.~\eqref{der_rel_box} and ~\eqref{der_rel_tt} from the Appendix and the expression for $F_c$ from Eq.~\eqref{fc_soln}, we get
\begin{align}
    F = \alpha_c \left(\Box F_c - \nabla^{t} \nabla_{t} F_c \right)
= 3 \alpha_c f_c e^{- \beta}\,. \label{f_soln}
\end{align}
At this point we can calculate the total energy density and pressure for the system. The spacetime we are considering
is Ricci flat (i.e., $R = 0$) and consequently from Eq.~\eqref{p_rho_tot_def} we have
\begin{align}
    \rho_{\text{Eul}} = - F, \,\,\,\ p_{\text{Eul}} = - n \frac{\partial F}{\partial n} + F. \label{p_rho_def_R=0}
\end{align}
From Eq.~\eqref{f_soln} we have
\begin{align}
  \rho_{\text{Eul}} = - F(r) = - 3\alpha_c f_c e^{- \beta(r)} = - 3 \alpha_c f_c \left(1 + \frac{k}{r} \right)^{1/2}. \label{rho_soln}
\end{align}
As we require $\rho_{\text{Eul}} \geq 0$, we must chose the constants $\alpha_c$ and $f_c$ such that
\begin{align}
    \alpha_c f_c <	0.
\end{align}
From Eqs.~\eqref{p_rho_def_R=0} and ~\eqref{fe_11} and using Eqs.~\eqref{der_rel_box} and ~\eqref{der_rel_rr} from the Appendix, the total pressure is
\begin{align}
    p_{\text{Eul}} = \alpha_c  \left(\Box F_c - \nabla^{r} \nabla_{r} F_c \right)    
    = \frac{1}{2} \alpha_c f_c \left[3\left(1 + \frac{k}{r}\right)^{1/2} + \left(1 + \frac{k}{r}\right)^{-1/2} \right]\label{p_soln}.
\end{align}
As we have assumed $\alpha_c f_c <	0$, the quantity $p_{\text{Eul}}$ is always a negative. From Eqs.~\eqref{rho_soln} and ~\eqref{p_soln} we can see that
\begin{align}
    \rho_{\text{Eul}} \rightarrow - 3 \alpha_c f_c \,\,\,\ \text{and} \,\,\,\ p_{\text{Eul}} \rightarrow 2 \alpha_c f_c, \,\,\,\ \text{when} \,\ r \rightarrow \infty.
\end{align}
So both the energy density $\rho_{\text{Eul}}$ and pressure $p_{\text{Eul}}$, reach constant values at a large distance.

From Eqs.~\eqref{p_soln} and ~\eqref{rho_soln} the EoS parameter in our case becomes:
\begin{align}
    w_{\text{Eul}} = \frac{p_{\text{Eul}}}{\rho_{\text{Eul}}} = - \frac{4r + 3k}{6 (r + k)}.
\end{align}
From the above expression of $w_{\text{Eul}}$, we see that $ w_{\text{Eul}} \rightarrow - {2}/{3}$ when $r \rightarrow \infty$ and it can also be verified that, although the energy density and pressure individually diverges as we approach the origin, the EoS parameter approaches a finite value $-1/2$. 

We can now calculate the particle number density. Using the chain rule of differentiation we can write 
$F - \left({n}/{n^{\prime}}\right) F^{\prime} = p_{\text{Eul}}$,
and then integrating the above equation we get:
\begin{align}
    n (r) = n_0 \hspace{1mm} \exp\left({\int dr \hspace{1mm} \frac{F^{\prime}}{F - p_{\text{Eul}}}} \right),
\end{align}
where $n_0$ is an integration constant. Using the expression for $F$ from Eq.~\eqref{f_soln} we can write the integrand as 
$$\frac{F^{\prime}}{F - p_{\text{Eul}}} = \frac{3 \alpha_c f_c \alpha^{\prime} e^{\alpha}}{3 \alpha_c f_c e^{\alpha} - \alpha_c f_c \left(2 + r \alpha^{\prime} \right) e^{\alpha}}\,,$$ 
which after some algebraic manipulations yields:
\begin{align}
    n(r) = n_0 \left(1 + \frac{3 k}{2r} \right). \label{n_soln}
\end{align}
This result predicts $n(r) \rightarrow n_0$ as $r \rightarrow \infty$, i.e., the particle number density reaches a constant value $n_0$, asymptotically when $r$ tends to a infinitely large value.
We can invert the relation in Eq.~\eqref{n_soln} to write $r$ as a function of $n$ as 
\begin{align}
    r = \frac{3k}{2} \left(\frac{n}{n_0} - 1 \right)^{-1}. \label{r(n)_soln}
\end{align}
Substituting this in Eq.~\eqref{f_soln} and Eq.~\eqref{fc(r)_soln} we can write $F$ and $F_c$ as a function of $n$ as follows:
\begin{align}
    F(n) &= \sqrt{3} \alpha_c f_c \left(1 + \frac{2n}{n_0} \right)^{1/2}, \label{F(n)_soln} \\
    F_c(n) &= \frac{3 f_c k^2}{4} \sqrt{\frac{2n + n_0}{3n_0}}\left[-\frac{3n_0(n-2n_0)}{2(n - n_0)^2} + \tanh^{-1}\left( \sqrt{\frac{3n_0}{2n + n_0}} \right) \right] + \widetilde{f_c}. \label{Fc(n)_soln}
\end{align}
The above relations complete the formal solution of the problem. 

Can the parameter $k$ be negative? The answer is a bit involved. If we assume the solutions hold till $r \to \infty$ then $k$ cannot be negative. If we assume $k$ as a negative number, i.e., $k = - |k|$,  then, from the expression of $n(r)$, we can see that, $n(r)< 0$ when $r < 3|k|/2$. As in this case the spacetime becomes  Schwarzschild, we expect a horizon at $r = |k|$,  then $n(r) < 0$, in the interval $|k| < r < 3 |k|/ 2$. As we can not allow the particle number density to assume negative values, we will only have to  consider positive $k$ values. 

On the other hand, if we take $k$ as a negative number, we can not have a Schwarzschild like solution which is physically consistent over all the possible values of the radial coordinate. In the case $k$ is negative, we have to simultaneously take the integration constant $n_0$ as a negative number, then the expression for the particle number density becomes
\begin{align}
    n (r) = \left|n_0 \right| \left(\frac{3 \left|k\right|}{2 r} - 1 \right)\,.
\end{align}
In this case, $n (r)$ is a positive valued function in the following range: $0 < r \leq {3 \left|k\right|}/{2}$. Consequently, in this case, we expect an event horizon at $r = \left|k\right|$. Although, in this case $n (r)$ remains positive within $\left|k\right| \leq r < 3\left|k\right|/2 $; it vanishes at $r = 3\left|k\right|/2$ and consequently one has to terminate the NMC scenario at $r = 3\left|k\right|/2$. One has to match some spacetime, for the outer region, with the one we have at $r = 3\left|k\right|/2$. 
However, in this case the expression for $r (n)$, $F (n)$ and $F_c (n)$ becomes:
\begin{align}
    r = \frac{3 \left|k\right|}{2} \left(1 + \frac{n}{\left|n_0\right|} \right)^{-1}
\end{align}
and 
\begin{align}
    F(n) &= \sqrt{3} \alpha_c f_c \left(1 - \frac{2n}{\left|n_0\right|} \right)^{1/2}, \label{F(n)_solnNegk} \\
    F_c(n) &= \frac{3 f_c k^2}{4} \sqrt{\frac{\left|n_0\right| - 2n}{3\left|n_0\right|}}\left[\frac{3 \left|n_0\right| \left(n + 2 \left|n_0\right| \right)}{2(n + \left|n_0\right|)^2} + \tanh^{-1}\left( \sqrt{\frac{3\left|n_0\right|}{\left|n_0\right| - 2n}} \right) \right] + \widetilde{f_c}. \label{Fc(n)_solnNegk}
\end{align}
Also, from the expressions of the metric components [Eq.~\eqref{met_form}], it can be seen that $g_{00}$ and $g_{11}$ change sign as one crosses the constant-$r$ hypersurface at $r = \left|k\right|$, towards the origin. So we consider only the region $\left|k\right| < r < 3 \left|k\right|/2$, within which, both the positivity of $n (r)$ and proper behavior of the metric components are ensured. 

From Eq.~\eqref{rho_soln}, we can see, that, the energy density $\rho_{\text{Eul}}$ becomes zero at $r = \left|k\right|$, and steadily increases to $\sqrt{3} \left|\alpha_c f_c \right|$ at $r = 3 \left|k\right|/2$. Similarly, from Eq.~\eqref{p_soln}, we can check, that the pressure $p_{\text{Eul}}$ becomes infinitely large and negative at $r = \left|k\right|$, and steadily decreases in magnitude to $ - \sqrt{3} \left|\alpha_c f_c \right|$ at $r = 3 \left|k\right|/2$. Consequently the EoS parameter $ w_{\text{Eul}}$ diverges to $- \infty$ at $r = \left|k\right|$ and steadily acquires the value $-1$ as one approaches $r = 3 \left|k\right|/2$.

As $k$ can be positive, our work shows that in the presence of curvature-fluid coupling one can accommodate matter just outside the Schwarzschild black hole event horizon, although this matter distribution cannot be extended to any arbitrary value of the radial coordinate. This feature may have interesting implications on the no-hair theorems related to black holes. From this result one cannot opine on the fate of the other kind of black holes, as the Kerr black hole and the Riessner-Nordstrom black hole. It will be interesting to see the fate of no-hair theorems in these cases in the presence of a curvature-fluid coupling.

In the present case, as we are dealing with a Ricci-flat spacetime, one can easily check that both the SETs agree in the present case and the above energy density and pressure are what will be used in a static fluid case. The results show that a standard vacuum solution in the minimally coupled scenario in GR can accommodate a particular kind of fluid which behaves like dark energy far away from the central singularity. This fluid exhibits negative pressure throughout. The alernative interpretation of this fact is more fascinating. If one has a vacuum solution in GR produced by the spacetime as discussed above and then introduce a fluid whose energy density and pressure are radial functions as given above, then this spacetime will accept this fluid at the cost of a NMC between curvature and fluid. Such negative pressure fluids can be produced in some dark energy models, in some epoch of cosmologiccal evolution. Moreover clustered dark energy can produce local patches of dark energy whose EoS may be modified due to local effects and in such cases one may have the above kind of spherically symmetric, static spacetime.

\section{\fontsize{12}{15}\selectfont Energy Conditions}
\label{econds}

\begin{table}[h!]
  \begin{center}
    \setlength{\tabcolsep}{24pt}
    \begin{tabular}{l c c} 
    \hline
    \hline
      Name & Statement & Conditions \\
      \hline
      Weak & $T_{\mu \nu} W^{\mu} W^{\nu} \geq 0$ & $\rho \geq 0$, $\rho + p_i \geq 0$\\ [1ex]
      Null & $T_{\mu \nu} k^{\mu} k^{\nu} \geq 0$ & $\rho + p_i \geq 0$\\ [1ex]
      Strong & $\left(T_{\mu\nu} - \frac{1}{2} T g_{\mu \nu} \right) U^{\mu} U^{\nu} \geq 0$ & $\rho + \sum_i p_i \geq 0$, $\rho + p_i \geq 0$ \\ [1ex]
      Dominant & $- T^{\mu}_{\nu} W^{\nu}$ future directed & $\rho \geq 0$, $\rho  \geq  |p_i| $ \\[1ex] 
      \hline
      \hline
    \end{tabular}
  \end{center}
  \caption{Summary of various energy conditions. Here $W^{\mu}$ is any timelike vector; $k^{\mu}$ is an arbitrary future-directed null vector and $U^{\mu}$ is any arbitrary normalized timelike vector.}
  \label{table:energy_conditions}
\end{table}
\begin{figure}
\begin{subfigure}{.5\textwidth}
  \centering
  \includegraphics[width=1.0\linewidth]{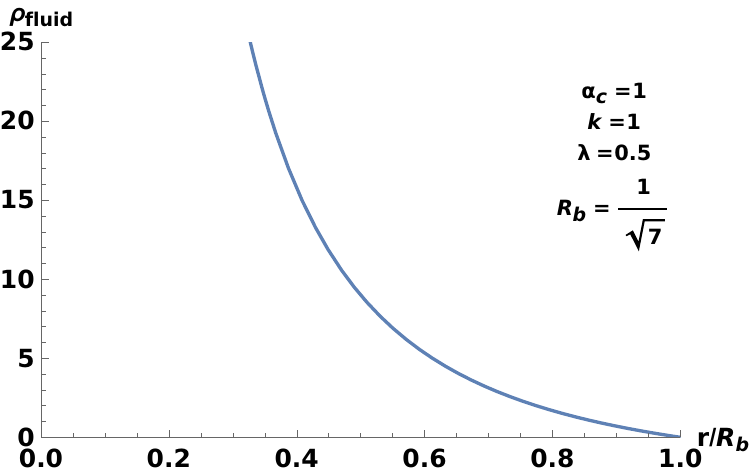}
  \caption{}
  \label{rhok1}
\end{subfigure}%
\begin{subfigure}{.5\textwidth}
  \centering
  \includegraphics[width=1.0\linewidth]{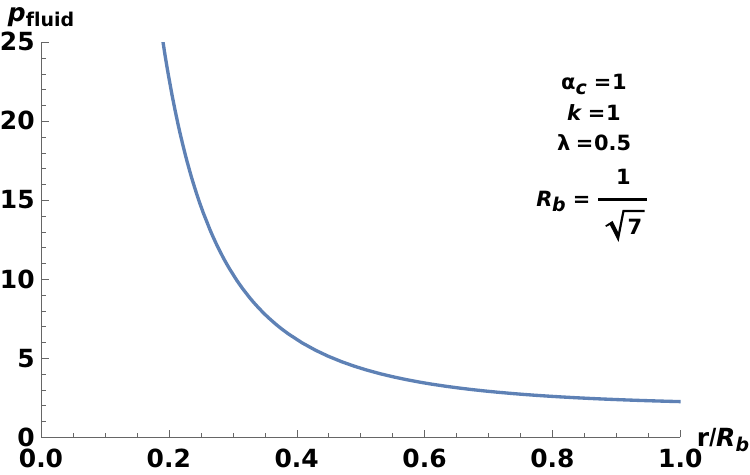}
  \caption{}
  \label{pk1}
\end{subfigure}%
\newline
\begin{subfigure}{.5\textwidth}
  \centering
  \includegraphics[width=1.0\linewidth]{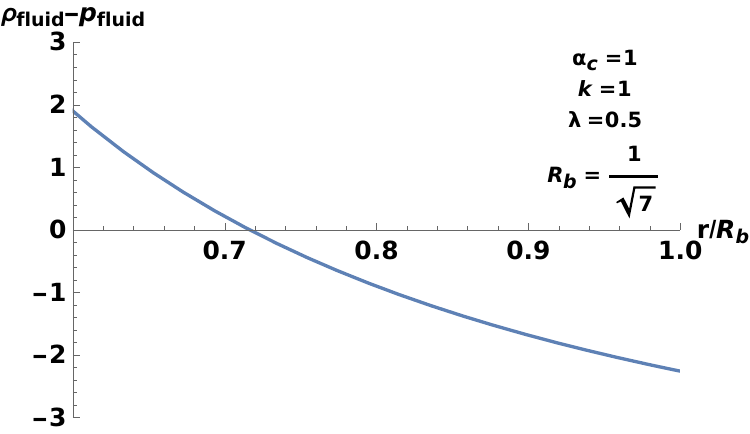}
  \caption{}
  \label{rhoFluid-pFluid}
\end{subfigure}%
\begin{subfigure}{.5\textwidth}
  \centering
  \includegraphics[width=1.0\linewidth]{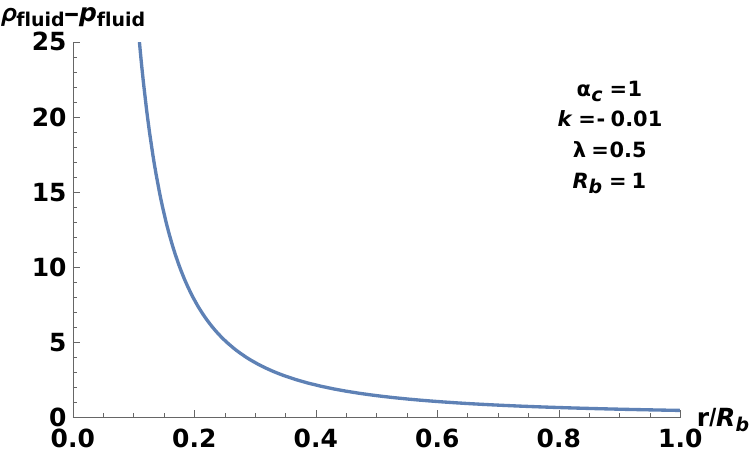}
  \caption{}
  \label{rhoFluid-pFluidNegk}
\end{subfigure}
\caption{In the above panel, we show the behavior of $\rho_{\text{Fluid}}$, $p_{\text{Fluid}}$ for the case of JMN-2 spacetime [Eq.~\eqref{pov1EC}]  as a function of radial coordinate, normalized with respect to $R_b$, in the context of various energy conditions. In Fig.~[\ref{rhok1}] and Fig.~[\ref{pk1}], we show the behavior of energy density and pressure, both of which are positive in the chosen interval. In Fig.~[\ref{rhoFluid-pFluid}], it is shown that, the pressure becomes greater in magnitude than the energy density after a certain value of radial coordinate, hence violating the dominant energy condition. All these considerations are drawn for positive value of the parameter $k$. However, for a suitably chosen negative value of $k$, all the energy energy conditions are satisfied. }
\label{}
\end{figure}

In Einstein's general relativity, there are several energy conditions [\cite{PoissonToolkit}, p. 219 of  Ref.~\cite{ WaldGR}] which are imposed on the stress-energy tensor which in turn put some physically reasonable conditions on the matter components. We summarize these energy conditions in the Table \ref{table:energy_conditions}. In our context, as because there are more than one possibility of  defining the stress-energy tensor (or the SET) and as a consequence we will like to analyze the various cases arising for the various kinds of SETs.
\begin{figure}
\begin{subfigure}{.5\textwidth}
  \centering
  \includegraphics[width=1.0\linewidth]{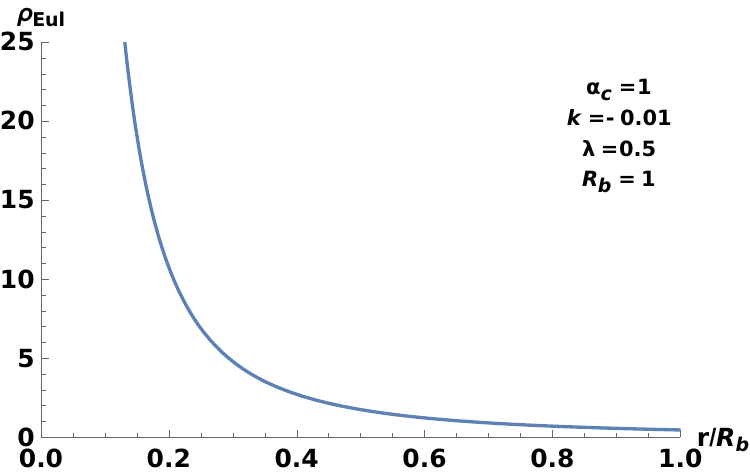}
  \caption{}
  \label{rhokneg}
\end{subfigure}%
\begin{subfigure}{.5\textwidth}
  \centering
  \includegraphics[width=1.0\linewidth]{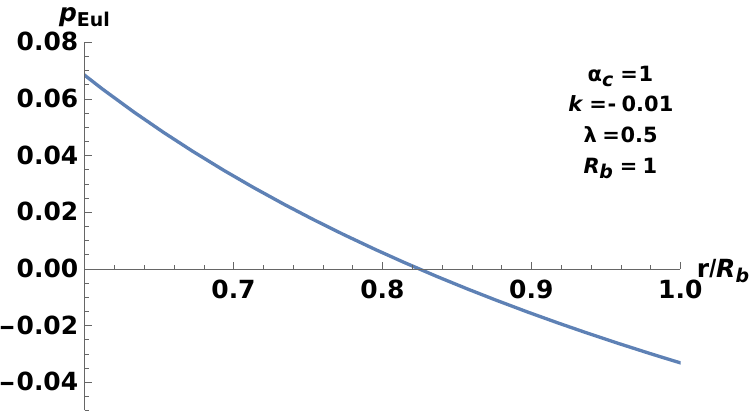}
  \caption{}
  \label{pkneg}
\end{subfigure}%
\newline
\begin{subfigure}{.5\textwidth}
  \centering
  \includegraphics[width=1.0\linewidth]{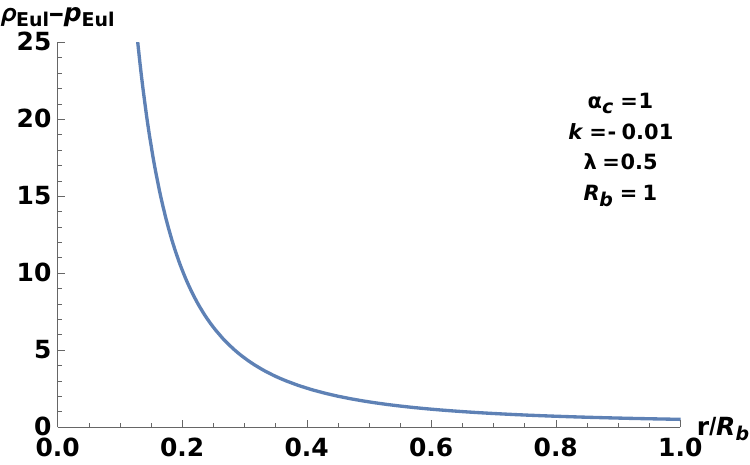}
  \caption{}
  \label{rhoEul-pEulNegk}
\end{subfigure}%
\begin{subfigure}{.5\textwidth}
  \centering
  \includegraphics[width=1.0\linewidth]{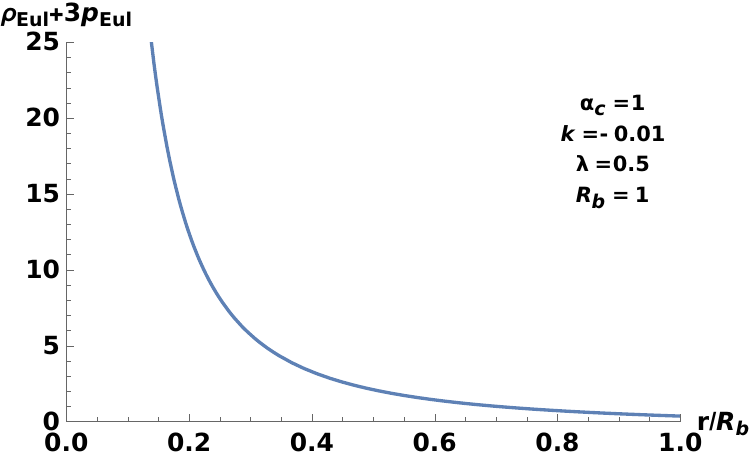}
  \caption{}
  \label{rhoEul+3pEulNegk}
\end{subfigure}%
\caption{In the above panel, we show the relevant behavior of $\rho_{\text{Eul}}$ and $p_{\text{Eul}}$ for some specific values of the parameters in the case of curvature-fluid coupling in JMN-2 spacetime. From Fig.~[\ref{rhoEul-pEulNegk}], we can see that, the difference $\rho_{\text{Eul}} - \left|p_{\text{Eul}} \right| $ is positive throughout the chosen interval. It can also be seen from Fig.~[\ref{rhoEul+3pEulNegk}], that, although the pressure becomes negative near $R_b$, the quantity $\left(\rho_{\text{Eul}} + 3 p_{\text{Eul}} \right)$ remains positive and consequently the strong energy condition holds. }
\label{}
\end{figure}

First we discuss the energy conditions related to the $C=0$ case in section \ref{c0cn0}. In this case one can easily see that $\rho_{\text{fluid}}$ and $p_{\text{fluid}}$ resembles a pure cosmological constant like behavior and hence the energy conditions here become identical to the energy conditions in the de Sitter spacetime. One can also discuss the case of the energy conditions in terms of $\rho_{\text{Eul}}$ and $p_{\text{Eul}}$. We do not present this discussion here as because we notice that $\rho_{\text{Eul}}=0$ here and this effective fluid will not satisfy the known  energy conditions. One must not be disturbed by the fact that $\rho_{\text{Eul}}$ and $p_{\text{Eul}}$ do not satisfy all the energy conditions as because these variables are effective variables.

Now we will consider the energy conditions in JMN-2 spacetime (the $C\ne 0$ case) which was obtained as a solution in section \ref{c0cn0}. In the first point of view [Eq.~\eqref{1stpov}] we have 
\begin{align}
   \rho_{\text{fluid}} =  \frac{1-\lambda ^2}{\left(2 - \lambda ^2 \right) \, r^2} - \frac{\alpha_c k \left(5 + \lambda^2 \right)}{2-\lambda^2}. \label{pov1EC}
\end{align}
We do not write the expression for the pressure due to it's complexity. From the above expression of the energy density, it can be seen that, in order to make it a positive valued function, we need to constrain the range of the radial coordinate in the following way:
\begin{align}
    0 < r \leq \sqrt{\frac{1 - \lambda^2}{\alpha_c k \left(5 + \lambda^2 \right)}}.
\end{align}
We can even identify the constant $R_b$ with this maximum value of the radial coordinate, i.e.,
\begin{align}
    R_b \equiv \left[\frac{1 - \lambda^2}{\alpha_c k \left(5 + \lambda^2 \right)} \right]^{\frac{1}{2}}.
\end{align}
If we plot $\rho_{\text{fluid}}$ and $p_{\text{fluid}}$  for a typical set of the constants; say $\lambda = \frac{1}{2}$, $\alpha_c = 1$ and $k = 1$, we can see that, both the energy density and pressure remain  positive in Figs.~[\ref{rhok1}, \ref{pk1}] up to $R_b$ and consequently, the weak, null and strong energy conditions are satisfied. The quantity $\left(\rho_{\text{fluid}} - p_{\text{fluid}}\right)$ becomes negative in Fig.~[\ref{rhoFluid-pFluid}] from a certain value of the radial coordinate up to $R_b$. So, except the dominant energy condition, which is not satisfied everywhere in the $r$-interval, all the other energy conditions are satisfied.

However, if we assume a negative value for the constant $k$, then for a typical set of values of parameters, $k = - 0.01$, $\lambda = 1/2$, $\alpha_c = 1$ and $R_b = 1$, the energy density $\rho_{\text{Fluid}}$ remains positive throughout $r \equiv (0, R_b)$. Although the pressure $p_{\text{Fluid}}$ becomes negative near $R_b$, we have $\left(\rho_{\text{Fluid}} + p_{\text{Fluid}} \right) > 0$, $\left(\rho_{\text{Fluid}} + 3p_{\text{Fluid}} \right) > 0$ and $\left(\rho_{\text{Fluid}} - \left|p_{\text{Fluid}} \right| \right) > 0$ [Fig.~\ref{rhoFluid-pFluidNegk}]. Consequently, in this case all energy conditions are satisfied. In this case we do not plot all the range of $r$ as we have previously mentioned that even in the presence of curvature-fluid coupling one cannot extend the JMN-2 spacetime to arbitrary high values of $r$ without hitting singularities. This spacetime has to be matched with a regular spacetime so that one has only one central singularity in this case.

In the second point of view [Eq.~\eqref{p_rho_tot_def}], we get an  expression for energy density given in Eq.~\eqref{rhosolnjmn2}. If we plot this function with respect to $r$, the function $\rho_{\text{Eul}}$ changes sign, and becomes negative, within the domain even for $r<R_b$. So, we do not get a well behaved energy density function. However, if we take the constant $k$ as a negative number, we can see from Eq.~\eqref{Fsol2}, that, $\rho_{\text{Eul}}$ is a positive valued function for every value of the radial coordinate. In this case, $R_b$, the value of $r$, where we want to truncate our metric solution [Eq.~\eqref{reverse_metric}], is a free parameter. In this case, we can plot the functions $\rho_{\text{Eul}}$ and $p_{\text{Eul}}$ for a typical set of values of the parameters, e.g., we have plotted for $\alpha_c  = 1$, $k =  - 0.01$ and $\lambda = \frac{1}{2}$, and found that, the energy density $\rho_{\text{Eul}}$ is positive throughout the interval $r \equiv (0, R_b =1)$ [Fig.~\ref{rhokneg}, \ref{pkneg}]. The pressure $p_{\text{Eul}}$ decreases from a diverging value from the origin towards $R_b$ and changes the sign after a certain value of the radial coordinate [see Fig.~[\ref{pkneg}] depending on the values of the parameters $\alpha_c$, $k$ and $\lambda$. In addition, we note the following from Fig.~[\ref{rhoEul-pEulNegk}] and  Fig.~[\ref{rhoEul+3pEulNegk}]:
\begin{align}
    \rho_{\text{Eul}} > 0, \,\,\, \rho_{\text{Eul}} + p_{\text{Eul}} > 0, \,\,\, \rho_{\text{Eul}} + 3 p_{\text{Eul}} > 0, \,\,\, \rho_{\text{Eul}} - p_{\text{Eul}} > 0, \,\,\, \rho_{\text{Eul}} - \left|p_{\text{Eul}} \right| > 0.
\end{align}
From the above relations, we conclude, that, all the energy conditions, viz., weak, strong, null and dominant are valid within the domain of the radial coordinate. 

Next we discuss the energy conditions in the case of the solution we studied in section \ref{unknfc}. In this case, for $m=1$, the spacetime solution had a vanishing value for the Ricci scalar and consequently in this case both the forms of the SETs discussed previously coincide. In the first or second points of view, we check whether the matter sector $T^{\left(\text{fluid}\right)}_{\mu\nu}$ [Eq.~\eqref{SET_fluid}] satisfies the energy conditions in the presence of curvature coupling. From Eq.~\eqref{rho_soln} and ~\eqref{p_soln} we write 
\begin{align}
  \rho_{\text{Eul}} =  3 \left|\alpha_c f_c \right| \left(1 + \frac{k}{r} \right)^{1/2}, \,\,\   
    p_{\text{Eul}} = - \frac{1}{2} \left|\alpha_c f_c \right| \left[3\left(1 + \frac{k}{r}\right)^{1/2} + \left(1 + \frac{k}{r}\right)^{-1/2} \right].
\end{align}
We recall that, in this case $R = 0$ and consequently, as stated earlier, we have:
\begin{align}
    \rho_{\text{Eul}} = \rho_{\text{fluid}}, \,\,\ \text{and} \,\,\ p_{\text{Eul}} = p_{\text{fluid}}. 
\end{align}
So, from the expressions for $\rho_{\text{Eul}}$ and $p_{\text{Eul}}$ we can write
\begin{align}
    \rho_{\text{fluid}} + p_{\text{fluid}} =  \frac{1}{2} \left|\alpha_c f_c \right| \left[3\left(1 + \frac{k}{r}\right)^{1/2} - \left(1 + \frac{k}{r}\right)^{-1/2} \right],
\end{align}
which is always positive for $0 < r < \infty$. So we conclude
\begin{align}
     \rho_{\text{fluid}} + p_{\text{fluid}} > 0.
\end{align}
As we have $\alpha_c f_c < 0$ we can write the above inequality as
\begin{align}
     \rho_{\text{fluid}} > \left| p_{\text{fluid}} \right|.
\end{align}
To check the strong energy condition let's consider the following expression:
\begin{align}
     \rho_{\text{fluid}} + 3 p_{\text{fluid}} = - \frac{3}{2} \left|\alpha_c f_c \right| \left[\left(1 + \frac{k}{r}\right)^{1/2} + \left(1 + \frac{k}{r}\right)^{-1/2} \right],
\end{align}
 which is always negative. So we conclude that, in the presence of the non-minimal coupling, all the energy conditions except the strong energy condition are satisfied by the fluid sector.
\section{Discussion and conclusion} 
\label{conclusion}

As discussed previously, there can be another way in which one can define an effective SET in the present case. In this scheme we  fix the gravitational constant as the Newtonian gravitational constant and define an effective stress-energy tensor with an attempt to make a correspondence with the Einstein's field equations. In this approach, we can write the field equations in Eq.~\eqref{f_e} in the form of Einstein's field equations [as in GR] with an effective stress-energy tensor, i.e.,
\begin{align}
    \frac{1}{\kappa}  G_{\mu \nu} =  \bar{T}^{\left(\text{eff}\right)}_{\mu \nu}, \label{f_e2}
\end{align}
with 
\begin{align}
\bar{T}^{\left(\text{eff}\right)}_{\mu \nu} \equiv g_{\mu \nu} \frac{F}{1 + \alpha_c F_c} - h_{\mu \nu} \left( \frac{\partial F}{\partial n} + \frac{\alpha_c}{2\kappa}  \frac{\partial F_c}{\partial n} R \right) \frac{n}{1 + \alpha_c F_c} - \frac{\alpha_c}{\kappa \left(1 + \alpha_c F_c \right)} \left( g_{\mu \nu} \Box F_c - \nabla_{\mu} \nabla_{\nu} F_c \right).
\end{align}
The above definition shows that the value of components of the effective SET will be same as the standard SET component values in GR. 

We can formally define the ``effective energy density" and ``effective pressures" as:
\begin{align}
    \bar{\rho}^{\left(\text{eff}\right)} \equiv - \left[\bar{T}^{ \left(\text{eff}\right)}\right]^0_0, \,\,\  \bar{p}^{\left(\text{eff}\right)}_1 \equiv \left[\bar{T}^{ \left(\text{eff}\right)}\right]^1_1, \,\,\ \bar{p}^{\left(\text{eff}\right)}_2 \equiv \left[\bar{T}^{ \left(\text{eff}\right)}\right]^2_2, \,\,\ \bar{p}^{\left(\text{eff}\right)}_3 \equiv \left[\bar{T}^{ \left(\text{eff}\right)}\right]^3_3. \label{p_rho_tot_def_2}
\end{align}
Here $\bar{p}^{\left(\text{eff}\right)}_i$'s are three principal pressures \cite{LSS}. Another way of defining the effective energy density and pressure is to imagine a hypothetical perfect fluid, whose effective stress-energy tensor is defined as
\begin{align}
    \frac{1}{\kappa}  G_{\mu \nu} =  \bar{T}^{\left(\text{eff}\right)}_{\mu \nu} = \bar{p}^{(\text{eff})} g_{\mu\nu} + \left(\bar{\rho}^{(\text{eff})} + \bar{p}^{(\text{eff})}\right) U_\mu U_\nu,
\end{align}
which can be inverted to write 
\begin{align}
    \bar{\rho}^{(\text{eff})} = U^\mu U^\nu \bar{T}^{\left(\text{eff}\right)}_{\mu \nu}, \,\,\,\,\,\, \bar{p}^{(\text{eff})} = \frac{1}{3} h^{\mu\nu} \bar{T}^{\left(\text{eff}\right)}_{\mu \nu}. \label{rho_p_eff_2.2} 
\end{align}
\begin{figure}
  \centering
  \includegraphics[width=0.5\linewidth]{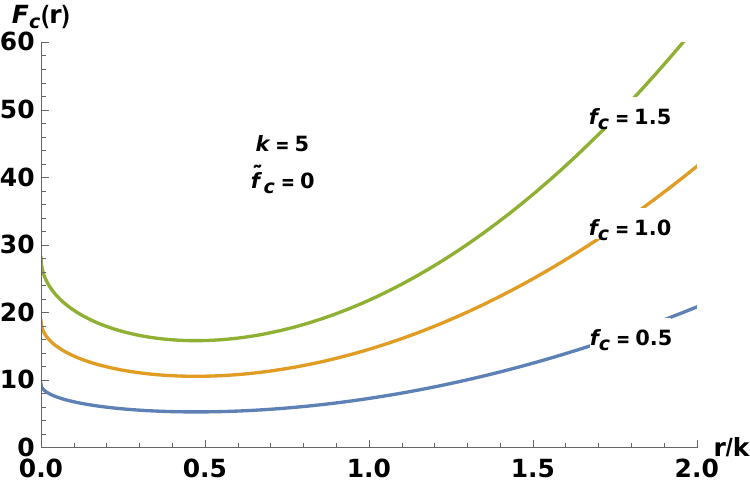}
\caption{ The above figure shows the variation of the conformal function $F_c (r)$ from Eq.~\eqref{fc(r)_soln} with the dimensionless variable $r/k$ for three different values of the integration constant $f_c$ with $\widetilde{f}_c = 0$. It can be seen, that, in general, the function $F_c (r)$ has a minima between $r = 0$ and $r = k$, and then, it increases steadily. Above three curves have been shown for a fixed value of the constant $k  = 5$. From the expression of $F_c (r)$, it can be understood that, increasing the value of $k$ will only increase the overall magnitude of $F_c$, which varies quadratically w.r.t. the constant $k$  [i.e., $F_c \sim k^2$], and consequently does not change the characteristics of the $F_c$ versus $r/k$ curve.}
\label{Fc(r)_vs_r}
\end{figure}
In this viewpoint the stress-energy tensor is conserved in the sense that, it's four-divergence vanishes, i.e.,
\begin{align}
    \nabla^{\mu} \bar{T}^{ \left(\text{eff}\right)}_{\mu \nu} = 0.
\end{align}
It can be seen that, $\bar{T}^{ \left(\text{eff}\right)}$ can't be written as a sum of the two parts where one corresponds solely to the fluid part and another corresponds to the conformal function as before. We rather have
\begin{align}
    \bar{T}^{ \left(\text{eff}\right)}_{\mu \nu} = \left(1 + \alpha_c F_c \right)^{-1} T^{ \left(\text{eff}\right)}_{\mu \nu}.
\end{align}
From the above discussion we see that we can in principle also define the SET as given above, but we avoided using it seriously for the following reasons. The first reason to avoid this approach is related to the fact that in this approach all the effects of the curvature-fluid coupling have been encoded purely in the effective fluid part of the modified field equation and the Einstein tensor part appears as it appears in GR. This fact makes this approach too asymmetric. The second reason for not seriously using this approach is related to the fact that the value of the SET components do remain the same as in GR in the minimally coupled case and consequently the value of the SET components do not show any NMC effect. There is another reason why we avoid this prescription and this reason is related to Ricci flat cases. In some of the cases we did consider  spacetimes for which the Einstein tensor vanishes and consequently from the above viewpoint the effective energy density and the pressures also vanished giving little information about the effect of NMC on the SET. For all these reasons we have not used this third viewpoint about the NMC system.

Another relevant question regarding the curvature-fluid coupling is related to the relative strength of the interaction. Can we approximate the effect of the NMC systems by their analogous minimally coupled solutions in GR? In other words, will the effect of the NMC be just like a perturbation over the minimally coupled solutions? For any reasonable $F(n)$, can one have a small value of $F_c(n)$ for all over spacetime for some small nonzero $\alpha_c$? If the answer is yes then we can use the minimally coupled solutions as rough approximations for the complicated NMC models. In the solutions which we have discussed, the answer turns out to be in the negative, we explain the situation by an example below. While treating the case related to $m=1$ in section \ref{unknfc} one might ask, under what conditions the conformal function $F_c(r)$ can be treated as a small correction to the Einstein theory.  From the field equations Eq.~\eqref{f_e}, we can see that, it will be well approximated by Einstein equations, in the minimally coupled case, when the following two conditions are satisfied:
\begin{align}
    \left|\alpha_c F_c\right| &\ll 1, \label{perturb_cond1}\\ \text{and} \,\,\, \left|\frac{\alpha_c}{2} h_{\mu \nu} n \frac{\partial F_c}{\partial n} R  - {\alpha_c} \left( g_{\mu \nu} \Box F_c - \nabla_{\mu} \nabla_{\nu} F_c \right) \right| &\ll \left| g_{\mu \nu} F - h_{\mu \nu} n\frac{\partial F}{\partial n} \right| \label{perturb_cond2}. 
\end{align}
As the functions $F(r)$ and $F_c(r)$ are not independent, rather connected through the field equations, it might not  be the case, where the above two conditions are simultaneously met for a given solution of the field equations.   Let us work out the first of the above conditions for the solution we have obtained. We can plot $F_c(r)$ for various set of values for $f_c$, as shown in   Fig.~[\ref{Fc(r)_vs_r}],  to observe the following generic feature: $F_c(r)$ first decreases with $r$,  reaching its minimum value somewhere between $r = 0$ and $r = k$ and then steadily increases with $r$. 
We can make a Taylor series expansion of $F_c(r)$ [Eq.~\eqref{fc(r)_soln} with $\widetilde{f}_c = 0$] around $r=0$ for $r <	k$ as
\begin{align}
    F_c (r) \approx \frac{3}{4} f_c k^2 -  \frac{3 f_c k^{2}}{4} \left(\frac{r}{k}\right)^{\frac{1}{2}} + \frac{f_c k^2}{4} \left(\frac{r}{k}\right) +  \frac{f_c k^{2}}{4}  \left(\frac{r}{k}\right)^{\frac{3}{2}} + \mathcal{O}\left(r^2\right).
\end{align}
In the above, $\mathcal{O}\left(r^2\right)$ represents the sum of all the terms which are second or higher order in $\left({r}/{k}\right)$. If  we keep only the first two terms and ignore all the other terms in the above series, the condition Eq.~\eqref{perturb_cond1} becomes
\begin{align}
    \frac{3}{4} \left|\alpha_c f_c \right| k^2 \left| 1 - \sqrt{\frac{r}{k}} \right| \ll 1.
\end{align}
We can simplify the above relation to write
\begin{align}
   1 + \frac{4}{3\left|\alpha_c f_c \right| k^2} \gg \sqrt{\frac{r}{k}} \gg 1 - \frac{4}{3\left|\alpha_c f_c \right| k^2}. 
\end{align}
It is now easy to understand that for a given value of $r/k$, we can always chose the values of $\alpha_c$ and the integration constant $f_c$  such that, the above relation is satisfied. This analysis shows that, at least, from the origin $r=0$ upto a certain range of the radial coordinate, we can chose the parameters of the theory such that the condition in Eq.~\eqref{perturb_cond1} is always satisfied.  Beyond that range of $r$, the function $F_c(r)$ increases indefinitely with $r$ and cannot anymore be treated as a small quantity. In addition, for this example, as all the components of the Einstein tensor vanishes, from Eq.~\eqref{secFE}, we can write  
\begin{align}
    \delta^{\mu}_{\nu} F - h^\mu_\nu  \frac{\partial F}{\partial n} n = - \left[\frac{\alpha_c}{2} h^\mu_\nu \frac{\partial F_c}{\partial n} n R  - {\alpha_c} \left( \delta^{\mu}_{\nu} \Box F_c - \nabla^{\mu} \nabla_{\nu} F_c \right) \right], \nonumber
\end{align}
and consequently both sides of Eq.~\eqref{perturb_cond2} are equal, 
which prevents one, to treat the contribution of the conformal function $F_c(r)$ as a perturbative correction. Generally, under no circumstances will the curvature-fluid coupled system produce a solution which is very close to the minimally coupled system in GR.
In some particular cases, for some particular region of space, one may have close resemblance of the NMC solution and the minimally coupled solution.

Next we discuss the case which we skipped while discussing the solutions in section \ref{unknfc}. For  $m = - 2$, in section \ref{unknfc}, for which the the metric components [Eq.~\eqref{metcompgenexam}] and the Ricci scalar [Eq.~\eqref{ricciscalargenexam}] become 
\begin{align}
    e^{ - 2\beta (r)} = e^{2 \alpha (r)} = 1 + k r^2, \,\,\, R = - 12k.
\end{align}
From Eq.~\eqref{ET_ex2}, the non-vanishing components of the Einstein tensor become
\begin{align}
    G^0_0 = G^1_1 = 
    G^2_2 = G^3_3 =  3k. \label{ET_2}
\end{align}
However, in this case, using Eqs.~\eqref{fe2_00} and ~\eqref{fe2_11}, we get the following solutions for $F (r)$ and $F_c (r)$:
 \begin{align}
     F_c (r) = \frac{C}{k} \sqrt{1 + k r^2} + \widetilde{C}, \,\,\,
    F (r) = 3 k \left(1 + \alpha_c \widetilde{C} \right) + 6 \alpha_c C \sqrt{1 + k r^2}. \label{Fsolm-2}
\end{align}
Here $C$ and $\widetilde{C}$ are integration constants. But, for this solutions, particle number density comes out as a constant. Since, the functions $F$ and $F_c$ were assumed to be functions of particle number density, this  is a contradiction. So, we conclude that, in this case, we do not get a consistent solution for the system.

In this context, we may ask, whether a pure Minkowski spacetime can sustain such a fluid which is conformally coupled with curvature. We can solve the field equations Eq.~\eqref{secFE} to get the following solutions as given in Ref.~\cite{DarkFluidKaushik}: $F = 6 C$ and $F_c = C r^2$, where $C$ is a constant. In this case, the particle number density comes out as a constant. As we have assumed that both  $F$ and the conformal coupling function $F_c$ are purely functions of the particle number density, then we will not be able to accommodate the previous results, and all the results given in the present paper are new. We hope the present work is logically more cogent.

In this paper we have tried to construct spherically symmetric, static spacetimes in the presence of curvature-fluid NMC. The actual 
solutions require various complex questions to be answered. In minimally coupled GR only the metric affects the SET of the matter component. In presence of curvature-fluid coupling, the first nontrivial question which comes up is related to the nature of the SET. In this case one has to first answer what will be called the source of curvature. This point becomes essential because curvature itself can now be directly related to the energy density and pressure of the source of curvature. The second cause of  complications is related to the strategy for obtaining the solutions. Apriori assuming a barotropic fluid with a known EoS will not give us the spacetime solution, as in the present case we must also have to specify the curvature-fluid coupling function $F_c(n)$. The modified field equations show that there are compatibility conditions between $F(n)$ and $F_c(n)$ if one knows the spacetime metric. On the other hand if one starts with some $F_c(n)$, one cannot choose  any arbitrary barotropic fluid to start with as in this case the nature of $F(n)$ will be dictated by the compatibility conditions stated above. The whole theory is formulated in such a way that $F(n)$ and $F_c(n)$ are assumed to be functions of particle number density $n(r)$ in a spherically symmetric, static spacetime. As a result, one assumes all the relevant functions in the modified field equations as functions of the radial coordinate $r$. This way of solving the system poses another complex question: Is $n(r)$ always invertible? If not, then one cannot rewrite $F(r)$ and $F_c(r)$ as $F(n)$ and $F_c(n)$. All these conceptual and technical questions make the spherically symmetric, static solutions in curvature-fluid coupled systems complex. Nonetheless, one does get interesting solutions in the present case and some of these solutions may have important astrophysical or cosmological properties. In this paper we have just touched the surface layer of the problems related to NMC. We hope to unravel more interesting results in the future when we delve deeper in this subject.    

\vskip 1cm
\centerline{\Large \bf Appendix}

\appendix
\section{Some results in spherically symmetric, static spacetimes}
\label{appA}
For convenience we note down the  non-zero components of the Christoffel symbols for the metric in Eq.~\eqref{SphSymmLine}
\begin{align}
    &\Gamma^{t}_{tr} = \alpha^{\prime}, &\Gamma^{r}_{tt} &=  e^{2\alpha -2\beta} \alpha^{\prime},  
    &\Gamma^{r}_{rr} &= \beta^{\prime}, \label{connection_coeffs}
    \\ \nonumber
    &\Gamma^{\vartheta}_{r\vartheta} =   \frac{1}{r},
    &\Gamma^{r}_{\vartheta\vartheta} &= - r e^{- 2\beta},
    &\Gamma^{\varphi}_{r\varphi} &= \frac{1}{r}, \\ \nonumber
    &\Gamma^{r}_{\varphi\varphi} = - r e^{- 2\beta}  \sin^2\vartheta, &\Gamma^{\vartheta}_{\varphi\varphi} &=
    - \sin\vartheta \cos\vartheta, &\Gamma^{\varphi}_{\vartheta\varphi} &=
    \frac{\cos\vartheta}{\sin\vartheta},  \\ \nonumber
\end{align}
and the expression of the Ricci scalar becomes
\begin{align}
    R = \frac{2}{r^2} - 2 e^{- 2\beta}\left\{{\alpha}^{\prime \prime} + {\alpha^{\prime}}^2 - \alpha^{\prime} \beta^{\prime} + \frac{2}{r} \left(\alpha^{\prime} - \beta^{\prime}\right) + \frac{1}{r^2} \right\}. \label{ricci_scalar_gen}
\end{align}
Here and throughout the paper, the prime on a function of radial coordinate $r$, designates a derivative with respect to the radial coordinate, $r$. For future reference we also note down that, in this spacetime
\begin{align}
    &\nabla^{t} \nabla_{t} f = e^{-2\beta} \alpha^{\prime} f^{\prime}, \label{der_rel_tt} \\
    &\nabla^r \nabla_r f = e^{-2\beta} \left(f^{\prime \prime} - \beta^{\prime} f^{\prime} \right), \label{der_rel_rr} \\ 
    &\nabla^{\vartheta} \nabla_{\vartheta} f = \frac{e^{-2\beta}}{r} f^{\prime} = \nabla^{\phi} \nabla_{\phi} f, \label{der_rel_thetatheta} \\ 
    &\Box f = e^{-2\beta} \left[f^{\prime \prime} + \left(\alpha^{\prime} - \beta^{\prime} + \frac{2}{r} \right) f^{\prime} \right], \label{der_rel_box}
\end{align}
for any scalar function $f$ which is a function of the radial coordinate $r$ only. 

For the generic static and  spherically symmetric spacetime having metric structure given in Eq.~\eqref{SphSymmLine}, the expression for the non-vanishing components of the Einstein tensor are:
\begin{align}
    &G^0_0 = e^{-2\beta} \left(\frac{1}{r^2} - \frac{2 \beta^{\prime}}{r} \right) - \frac{1}{r^2},  \\
    &G^1_1 = e^{-2\beta} \left(\frac{1}{r^2} + \frac{2 \alpha^{\prime}}{r} \right) - \frac{1}{r^2}, \\
    &G^2_2 = G^3_3 = e^{-2\beta} \left(\alpha^{\prime \prime} + {\alpha^{\prime}}^2 - \alpha^{\prime} \beta^{\prime}  + \frac{\alpha^{\prime} -\beta^{\prime}}{r} \right).  \label{ETgen}
\end{align}
For the particular forms of the metric components given in Eq.~\eqref{metcompgenexam}, the above formulae yields
\begin{align}
        G^0_0 = - \frac{ (m - 1) k}{r^{m+2}}, \,\,\,  
    G^1_1 = - \frac{(m - 1) k}{r^{m+2}}, \,\,\,
    G^2_2 = G^3_3 =  \frac{m (m - 1) k}{2}\frac{1}{r^{m+2}}. \label{ET_ex2}
\end{align}
For the spacetime metric solution in subsection (3.1.1), the non-vanishing components  of the Einstein tensor are the following:
\begin{align}
      G^0_0 = \frac{e^{-2 \beta} - 1}{r^2}, \,\,\, G^1_1 = \frac{e^{ - 2\beta} \left(2 \widetilde{\beta} + 1 \right) - 1}{r^2}, \,\,\, G^2_2 = G^3_3 = \frac{e^{- 2\beta} {\widetilde{\beta}}^2}{r^2}. \label{ET_reverse_metric_Czero}
\end{align}


 The constants [$\widetilde{P}$'s and $P$'s] in the  expression for $p_{\text{Eul}}$, which is given in Eq.~\eqref{pSolCzeroPov3}  are given by
\begin{align}
    \widetilde{P}_0 &\equiv \left(e^{- 2 \beta} - 1 \right) \left( \widetilde{\mathcal{N}_1} + 2 \mathcal{N}_1 \right), \nonumber \\   \widetilde{P}_1 &\equiv \alpha_c \left[1 - e^{-2\beta} \left(1 + \widetilde{\beta} + {\widetilde{\beta}}^2 \right) \right] \left( \widetilde{\mathcal{N}_1} - 2 \mathcal{N}_1 \right) + \left(e^{- 2 \beta} - 1 \right) \left( \widetilde{\mathcal{N}_2} + 2 \mathcal{N}_2 \right), \nonumber \\
   \widetilde{P}_2 &\equiv \alpha_c \left[1 - e^{-2\beta} \left(1 + \widetilde{\beta} + {\widetilde{\beta}}^2 \right) \right] \left( \widetilde{\mathcal{N}_2} - 2 \mathcal{N}_2 \right), \nonumber \\
   P_0 &\equiv \widetilde{\mathcal{N}_1}, \nonumber \\
   P_1 &=  \widetilde{\mathcal{N}_2}. \label{P's}
\end{align}

The non-vanishing components of the Einstein tensor in JMN-2 spacetime [Eq.~\eqref{JMN2metric}] are:
\begin{align}
    G^0_0 = - \frac{1-\lambda ^2}{\left(2 - \lambda ^2 \right) \, r^2},  \,\,\,\,\,\,\
    G^1_1 = G^2_2 = G^3_3 = \frac{\left(1- \lambda ^2\right)^2 \left[1-\left(\frac{r}{R_b}\right)^{2 \lambda }\right]}{\left(2 -\lambda ^2\right) r^2 \left[(1 + \lambda )^2 -(1 - \lambda)^2 \left(\frac{r}{R_b}\right)^{2 \lambda }\right]}. \label{ET_jmn2}
\end{align}



\end{document}